\def\answ{b }
\input harvmac
\input labeldefs.tmp
\writedefs
\overfullrule=0pt

% redefine figures ... 
\input epsf
\def\fig#1#2#3{
\xdef#1{\the\figno}
\writedef{#1\leftbracket \the\figno}
\nobreak
\par\begingroup\parindent=0pt\leftskip=1cm\rightskip=1cm\parindent=0pt
\baselineskip=11pt
\midinsert
\centerline{#3}
\vskip 12pt
{\bf Fig.\ \the\figno:} #2\par
\endinsert\endgroup\par
\goodbreak
\global\advance\figno by1
}
\newwrite\tfile\global\newcount\tabno \global\tabno=1
\def\tab#1#2#3{
\xdef#1{\the\tabno}
\writedef{#1\leftbracket \the\tabno}
\nobreak
\par\begingroup\parindent=0pt\leftskip=1cm\rightskip=1cm\parindent=0pt
\baselineskip=11pt
\midinsert
\centerline{#3}
\vskip 12pt
{\bf Tab.\ \the\tabno:} #2\par
\endinsert\endgroup\par
\goodbreak
\global\advance\tabno by1
}
\def\der{\partial}
\def\d{{\rm d}}
\def\e#1{{\rm e}^{#1}}

\font\cmss=cmss10 \font\cmsss=cmss10 at 7pt
\def\R{\relax{\rm I\kern-.18em R}}
\def\Z{\relax\ifmmode\mathchoice
{\hbox{\cmss Z\kern-.4em Z}}{\hbox{\cmss Z\kern-.4em Z}}
{\lower.9pt\hbox{\cmsss Z\kern-.4em Z}}
{\lower1.2pt\hbox{\cmsss Z\kern-.4em Z}}\else{\cmss Z\kern-.4em Z}\fi}
%

 % careful... \Im, \im, \Imm
\def\Gam{{\mit\Gamma}}

\def\Th{Thistlethwaite}
%
%%%%%%%%%%%%%%%%%%%%%%%
% preprint # in refs ?
\def\pre#1{ (preprint {\tt #1})}%use this to give preprint # in refs
%\def\pre#1{}%use this NOT to give preprint # in refs
%%%%%%%%%%%%%%%%%%%%%%%%%%%%%%%%%%%%%%%%%%%%%%%%%%%%%%%%%%%%
%
%References
%
%\lref\XXX{N.N. XXX, {\it title }, .}
% \refs{\..{--}\...}
\lref\Hash{R.~Sedgewick,
{\sl Algorithms in C} (Addison-Wesley, 1990).}
\lref\Takacs{L. Tak\'acs,
{\sl Enumeration of rooted trees and forests},
{\it Math. Scientist} {\bf 18}, 1--10 (1993).}
\lref\Liskovets{V.A. Liskovets, {\sl Sequence A054499} in
N.J.A. Sloane (red.), {\sl The On-Line Encyclopedia of Integer Sequences},
published electronically at 
http://www.research.att.com/\~\/njas/sequences/.}
\lref\AV{I.Ya. Arefeva and I.V. Volovich, 
{\sl Knots and Matrix Models}, {\it Infinite Dim.
Anal. Quantum Prob.} {\bf 1} (1998) 1\pre{hep-th/9706146}).}
\lref\BIPZ{E. Br{\'e}zin, C. Itzykson, G. Parisi and J.-B. Zuber, 
{\sl Planar Diagrams}, {\it Commun. Math. Phys.} {\bf 59} (1978) 35--51.}
\lref\BIZ{D. Bessis, C. Itzykson and J.-B. Zuber, 
{\sl Quantum Field Theory Techniques in Graphical Enumeration},
{\it Adv. Appl. Math.} {\bf 1} (1980) 109--157.}
\lref\DFGZJ{P. Di Francesco, P. Ginsparg and J. Zinn-Justin, 
{\sl 2D Gravity and Random Matrices, }{\it Phys. Rep.} {\bf 254} (1995)
1--133.}
\lref\tH{G. 't Hooft, 
{\sl A Planar Diagram Theory for Strong 
Interactions}, {\it Nucl. Phys.} {\bf B 72} (1974) 461--473.}
\lref\HTW{J. Hoste, M. Thistlethwaite and J. Weeks, 
{\sl The First 1,701,936 Knots}, {\it The Mathematical Intelligencer}
{\bf 20} (1998) 33--48.}
\lref\KPZ{V.~G.~Knizhnik, A.~M.~Polyakov and A.~B.~Zamolodchikov,
{\sl Fractal structure of 2D quantum gravity},
{\it Mod.~Phys.~Lett.~A} {\bf 3}, 819--826 (1988);
F.~David,
{\sl Conformal field theories coupled to 2D gravity in the conformal gauge},
{\it Mod.~Phys.~Lett.~A} {\bf 3}, 1651--1656 (1988);
J.~Distler and H.~Kawai,
{\sl Conformal field theory and 2D quantum gravity},
{\it Nucl.~Phys.} {\bf B 321}, 509 (1989).}
\lref\MTh{W.W. Menasco and M.B. \Th, 
{\sl The Tait Flyping Conjecture}, {\it Bull. Amer. Math. Soc.} {\bf 25}
(1991) 403--412; 
{\sl The Classification of Alternating 
Links}, {\it Ann. Math.} {\bf 138} (1993) 113--171.}
\lref\Ro{D. Rolfsen, {\sl Knots and Links}, Publish or Perish, Berkeley 1976.}
\lref\STh{C. Sundberg and M. Thistlethwaite, 
{\sl The rate of Growth of the Number of Prime Alternating Links and 
Tangles}, {\it Pac. J. Math.} {\bf 182} (1998) 329--358.}
\lref\Tutte{W.T. Tutte, {\sl A Census of Planar Maps}, 
{\it Can. J. Math.} {\bf 15} (1963) 249--271.}
\lref\Zv{A. Zvonkin, {\sl Matrix Integrals and Map Enumeration: An Accessible
Introduction},
{\it Math. Comp. Modelling} {\bf 26} (1997) 281--304.}
\lref\KM{V.A.~Kazakov and A.A.~Migdal, {\sl Recent progress in the
theory of non-critical strings}, {\it Nucl. Phys.} {\bf B 311} (1988)
171--190.}
\lref\KP{V.A.~Kazakov and P.~Zinn-Justin, {\sl Two-Matrix Model with
$ABAB$ Interaction}, {\it Nucl. Phys.} {\bf B 546} (1999) 647
\pre{hep-th/9808043}.}
\lref\ZJZ{P.~Zinn-Justin and J.-B.~Zuber, {\sl Matrix Integrals
and the Counting of Tangles and Links},
to appear in the proceedings of the 11th 
International Conference on Formal Power Series and Algebraic 
Combinatorics, Barcelona June 1999
\pre{math-ph/9904019}.}
\lref\ZJZb{P.~Zinn-Justin and J.-B.~Zuber, {\sl On the Counting of Colored
Tangles}, {\it Journal of Knot Theory and its Ramifications} 
9 (2000) 1127--1141\pre{math-ph/0002020}.}
\lref\PZJ{P.~Zinn-Justin, {\sl Some Matrix Integrals
related to Knots and Links}, proceedings
of the 1999 semester of the MSRI ``Random Matrices
and their Applications'', MSRI Publications Vol. 40 (2001)
\pre{math-ph/9910010}.}
\lref\PZJb{P.~Zinn-Justin, {\sl The Six-Vertex Model on
Random Lattices}, {\it Europhys. Lett.} 50 (2000) 15--21\pre{cond-mat/9909250}.}
\lref\IK{I.~Kostov, {\sl Exact solution of the Six-Vertex
Model on a Random Lattice}, {\it Nucl. Phys.} {\bf B 575} (2000) 513-534\pre{hep-th/9911023}.}
\lref\KAUF{L.H.~Kauffman, {\sl Knots and physics},
World Scientific Pub Co (1994).}
\nref\Lorentz{P.~Di Francesco, E.~Guitter and C.~Kristjansen,
{\sl Integrable 2D Lorentzian Gravity and Random Walks},
{\it Nucl.~Phys.} {\bf B 567} (2000) 515--553\pre{hep-th/9907084}.}
\nref\Jensen{I.~Jensen,
{\sl Enumerations of Plane Meanders} \pre{cond-mat/9910313};
{\sl A Transfer Matrix Approach to the Enumeration of Plane Meanders},
{\it J.~Phys.~A}, to appear\pre{cond-mat/0008178}.}
\nref\Meanders{P.~Di Francesco, E.~Guitter and J.L.~Jacobsen,
{\sl Exact Meander Asymptotics: a Numerical Check},
{\it Nucl.~Phys.} {\bf B 580} (2000) 757--795\pre{cond-mat/0003008}.}
\lref\JZJ{J.~L.~Jacobsen and P.~Zinn-Justin, {\sl A Transfer Matrix approach to the Enumeration of Knots}\pre{math-ph/0102015}.}
\lref\PZJc{P.~Zinn-Justin, {\sl The General O(n) Quartic Matrix Model and its application to Counting Tangles and Links}\pre{math-ph/0106005}.}
\lref\JZJc{J.~L.~Jacobsen and P.~Zinn-Justin, work in progress.}
\lref\KoS{I.K.~Kostov, {\it Mod. Phys. Lett.} A4 (1989), 217\semi
M.~Gaudin and I.K.~Kostov, {\it Phys. Lett.} B220 (1989), 200\semi
I.K.~Kostov and M.~Staudacher, {\it Nucl. Phys.} {\bf B 384} (1992), 459.}
%%%%%%%%%%%%%%%%%%%%%%%%%%%%%%%%%%%%%%%%%%%%%%%%%%%%%%%%%%%%%%%%%%%%%%
\Title{
\vbox{\baselineskip12pt\hbox{\tt math-ph/0104009}}}
{{\vbox {
\vskip-10mm
\centerline{A Transfer Matrix approach}
\vskip2pt
\centerline{to the Enumeration of Colored Links}
}}}
\medskip
\centerline{J.~L.~Jacobsen {\it and} P.~Zinn-Justin}\medskip
\centerline{\sl Laboratoire de Physique Th\'eorique et Mod\`eles Statistiques}
\centerline{\sl Universit\'e Paris-Sud, B\^atiment 100}
\centerline{\sl 91405 Orsay Cedex, France}
\vskip .2in
% abstract
\noindent 

We propose a transfer matrix algorithm for the enumeration of alternating
link and tangle diagrams, giving a weight $n$ to each
connected component. Considering more general tetravalent diagrams
with self-intersections and tangencies allows us to treat topological
(flype) equivalences. This is done by means of a finite
renormalization scheme for an associated matrix model.  We
give results, expressed as polynomials in $n$, for the various
generating functions up to order 19 (2-legged tangle diagrams), 15 
(4-legged tangles) and 11 (6-legged tangles) crossings.  The limit
$n\to\infty$ is solved explicitly. We then analyze the large-order
asymptotics of the generating functions. For $0\le n \le 2$ good agreement
is found with a conjecture for the critical exponent, based on the KPZ
relation.

\Date{04/2000}
%\draft

\newsec{Introduction}

It is well-known that a $d$-dimensional system in statistical mechanics can be
conceived as a $(d-1)$-dimensional quantum field theory, by distinguishing one
of the spatial coordinates as the direction of time. This correspondence lies
at the heart of the {\it transfer matrix} formalism, where a linear operator
is used to describe the discrete time evolution of the corresponding quantum
system. More generally, transfer matrices have numerous applications for the
combinatorial enumeration of discrete objects for which a definite direction
(the transfer, or time, direction) can be singled out. Recently, this
combinatorial aspect has come into focus through the enumeration of various
objects pertaining to two-dimensional quantum gravity
\refs{\Lorentz,\Jensen,\Meanders,\JZJ}, such as plane meanders. Common to
these examples is the existence of a preferred direction (e.g.\ the river, in
the case of meanders), which can be straightforwardly promoted to the time
direction.

In a previous paper \JZJ\ we have shown how this scheme also applies to the
enumeration of alternating knot diagrams. Here the knot itself defines the
transfer direction, since the algorithm essentially consists in reading the
knot starting from one ``ingoing'' leg and ending at the other ``outgoing''
leg.

It is however far from obvious how this principle may generalize to the case
of {\it link diagrams} with more than one connected component. This is the
purpose of the present paper. More specifically, the final goal is to count
alternating tangles {\it at fixed number of connected components}; it is
therefore a generalization of the counting of alternating tangles
with minimum number of components done in \JZJ,
but also of the counting of
alternating tangles of \refs{\STh,\ZJZ} and of oriented alternating tangles
of \ZJZb. We also present some results for tangles with a higher number of
outgoing strings (``external legs''), instead of just four as
in the publications mentioned above.
We shall in what follows present not
just one, but two rather different transfer matrices addressing this
enumeration problem.

After the definitions in section 2, which include various intermediate
generating functions needed in the calculation, we shall present in section 3
the basic ideas behind the two proposed transfer matrices for alternating tangle
diagrams. Section 4 is devoted to more technical details on the actual
implementation of these ideas on a computer, and section 5 gives the numerical
results and their analysis. Finally, in Section 6, we discuss how our
algorithms may be adapted to various other problems of interest in graph
theory and statistical physics.

\newsec{Definitions of the generating functions}

The objects we want to consider are tangles with $2k$ ``external legs'',
that is roughly speaking the data 
of $k$ intervals embedded in a ball $B$ and whose endpoints are
given distinct points on the boundary $\der B$, plus an arbitrary number of
(unoriented) circles
embedded in $B$, all intertwined, and considered up to orientation
preserving homeomorphisms of $B$ that reduce to the identity on $\der B$.
Tangles with $4$ external legs will be simply called tangles.
The rest of the basic definitions is
identical to those given in \JZJ. We represent these objects using diagrams,
and restrict ourselves to alternating diagrams. This implies in particular
that tangles can be considered as flype equivalence classes of diagrams \MTh.

Our goal is to count the number of prime tangles with a certain number of
external legs and connected components.
We shall relate in this section their generating functions to a
simpler, more directly computable quantity, which is the following triple
generating function 
\eqn\gen{
G(n,g_1,g_2)=\sum_{k,p_1,p_2=0}^\infty a_{k,p_1,p_2} n^k g_1^{p_1}
g_2^{p_2}}
where $a_{k,p_1,p_2}$ is the number of topologically
inequivalent open curves in the plane going from $(-\infty,0)$ to
$(+\infty,0)$ together with $k$ circles, connected together by $p_1$
regular intersections and $p_2$ tangencies, see Fig.~\linktang. Here
$n$, $g_1$ and $g_2$ are formal parameters, which can be evaluated at
arbitrary complex values; however, it is natural to identify $n$ with
a number of {\it colors} one can assign to any of the closed loops of
the diagram, so that the factor $n^k$ correctly counts the total
number of possible colorings of the diagram (assuming the external
legs to carry a fixed color). The coefficient $a_{k,p,0}$ of the
double generating function $G(n,g_1=g,g_2=0)$ possesses the following
interpretation: it is the number of alternating tangle diagrams with $2$
external legs, $k$ circles (i.e.\ $k+1$ connected components) and $p$
crossings. The general coefficients do not possess such a clear
knot-theoretic interpretation; however, they are needed to take into
account the flyping equivalence (see \refs{\PZJ,\PZJc}).
\fig\linktang{Open curve and a circle with intersections (green dots) and tangencies (red dots).}{\epsfxsize=5.5cm\epsfbox{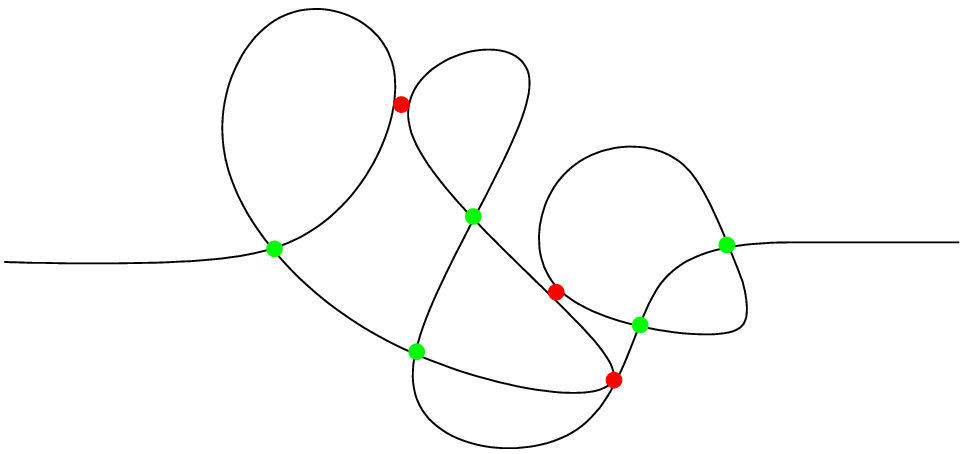}} 

\fig\types{Tangles of types 1 and 2 are distinguished by the two ways of
connecting their external legs.}
{\epsfxsize=8cm\epsfbox{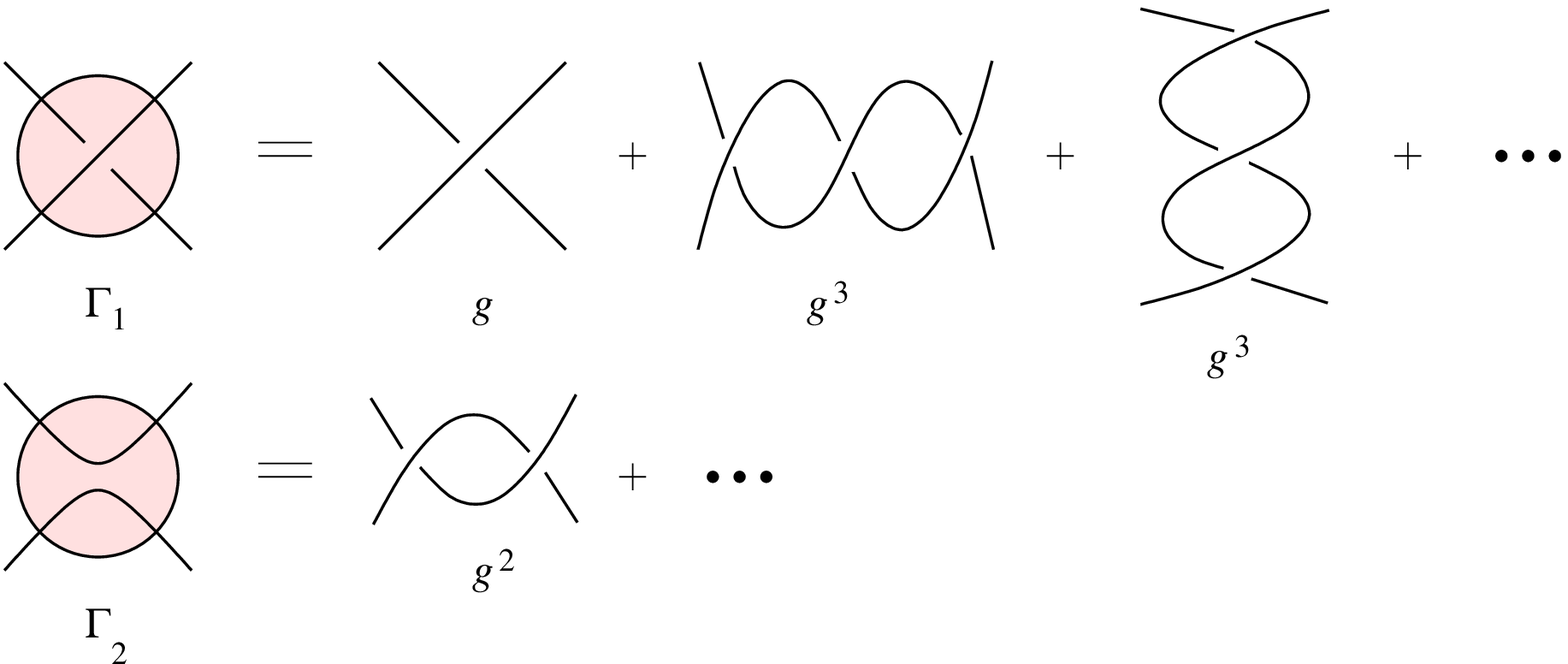}}
Next we define the generating functions of tangles $\Gam_1$ and
$\Gam_2$ (see Fig.~\types), which are necessary for the flyping equivalence.
They are given by:
\eqna\gend
$$\eqalignno{
G&=1+g_1 F_1+ g_2 F_2&\gend a\cr
{\der\over\der g_2} F_1&={\der\over\der g_1} F_2&\gend b\cr
4F_1&=nG_1+2G_2&\gend c\cr
2F_2&=G_1+(n+1)G_2&\gend d\cr
\Gam_1&=G_1&\gend e\cr
\Gam_2&=G_2 - G^2&\gend f\cr
}$$
via intermediate functions $F_1$, $F_2$, $G_1$, $G_2$.
Note that inverting Eqs.~\gend{c,d} requires $n\ne 1,-2$.
These special values of $n$ will be investigated in detail in \PZJc.

As in \JZJ, we introduce an extra parameter to count edges of the diagram,
according to the following definitions: $G(n,g_1,g_2,t)\equiv{1\over t}
G(n,g_1/t^2,g_2/t^2)$ and $\Gam_i(n,g_1,g_2,t)\equiv{1\over t^2}
\Gam_i(n,g_1/t^2,g_2/t^2)$.

The parameters $t$, $g_1$ and $g_2$ must then 
be chosen as a function of $n$ and $g$ according
to the following {\it renormalization procedure} (see \PZJc):
\eqna\ren
$$\eqalignno{
1&=G(n,g_1(n,g),g_2(n,g),t(n,g))&\ren a\cr
g_1(n,g)&=g(1-2H'_2(n,g))&\ren b\cr
g_2(n,g)&=-g(H'_1(n,g)+V'_2(n,g))&\ren c\cr
}
$$
where $H'_1(n,g)$, $H'_2(n,g)$ and $V'_2(n,g)$ are auxiliary quantities defined by:
\eqna\gene
$$\eqalignno{
H'_2\pm H'_1&={(1\mp g)(\Gam_2\pm\Gam_1)\mp g
\over 1+(1\mp g)(\Gam_2\pm\Gam_1)\mp g}&\gene a\cr
H'_2+H'_1+nV_2&={(1- g)(\Gam_2+(n+1)\Gam_1)- g
\over 1+(1- g)(\Gam_2+(n+1)\Gam_1)- g}&\gene b\cr
}$$
These equations are independent only for $n\ne 0$, but have a smooth $n\to 0$
limit which is given in \JZJ.

$\Gam_1(n,g_1(n,g),g_2(n,g),t(n,g))$ and
$\Gam_2(n,g_1(n,g), g_2(n,g),t(n,g))$,
once equations \ren{}\ are solved, 
are the desired generating functions for
the number of prime alternating tangles of types 1 and 2 respectively \ZJZb\
(see Fig.~\types). The total number of tangles is given by $\Gam_1+2\Gam_2$.
Similarly, one can define more general generating functions in the variables
$n$, $g_1$ and $g_2$ which count tangles with more external legs; an explicit
example will be given in Section 5.

\newsec{The transfer matrices for alternating tangles}
We now turn to the description of the two transfer matrices.
The basic idea is common to both of them:
starting from an initial state consisting of {\it all} external legs 
(i.e.\ two in the case under consideration),
the system is time evolved through the addition of $n$ intersections,
until an empty final state is obtained. 

Supposing the tangle diagram oriented from left to right, the first
algorithm procedes by always evolving the uppermost vertex.
We describe the details of this ``single-step'' algorithm in Section 3.1.
The second algorithm, on the contrary, evolves all parts of the diagram
simultaneously, adding one vertex to each of them in a given time step.
In this way, the time can be defined as the geodesic distance from the
pair of external legs. The details of this ``geodesic'' algorithm can
be found in Section 3.2.

As in \JZJ, we shall first concentrate on the enumeration of (prime,
alternating) tangle {\it diagrams} with two external legs, which are related to
the generating function $G(n,g=g_1,g_2=0)$ of Section 2. Adding tangencies,
which is needed to take into account the flyping equivalence, will be
discussed in Section 3.3, since it is an elementary extension of the
algorithms.

Another minor modification of the algorithms will enable us to enumerate
diagrams with more than two external legs; we shall develop this point in
Section 3.4.

In both algorithms, the needs for CPU-time and memory increase exponentially
with the system size $p$, though mercifully much more slowly than the number
of knot diagrams actually being enumerated. As will become clear shortly, the
single-step algorithm favors speed at the expense of memory consumption, while
for the geodesic algorithm it is the other way around. However, since in
practice both of these parameters are limiting factors for the maximally
obtainable system size, it is a priori not clear which of the algorithms
performs best. We defer a detailed comparison to Section 4.3, and it turns out
that the single-step algorithm comes out as the winner. Incidentally, even in
the case of knots (one connected component), it performs slightly better than
the algorithm described in \JZJ.

\subsec{The single-step algorithm} Let us briefly recall the working principle
behind the knot enumeration algorithm presented in \JZJ. Reading the
two-legged knot diagram from the first ``ingoing'' to the second ``outgoing''
leg, and calling at any instant the edge being read the ``active line'', there
are two possibilities at each time step: 1) The active line is crossed by a
line segment with edge labels that have not previously been encountered. We
then add the new line segment to the current state. 2) The active line is
identified with one of the endpoints of a line segment previously encountered.
We then join the active line to that endpoint, wind around the line segment in
question, and identify the new position of the active line with the opposite
endpoint.

\fig\singlestep{Working principle of the single-step algorithm.
a) A two-legged knot diagram with $p=6$ intersections and $k=3$
connected components. The edges are labelled from A to M.
b) The same diagram in the time-slice representation.
For reasons of clarity, the time slices are not drawn in chronological
order.}
{\epsfxsize=14cm\epsfbox{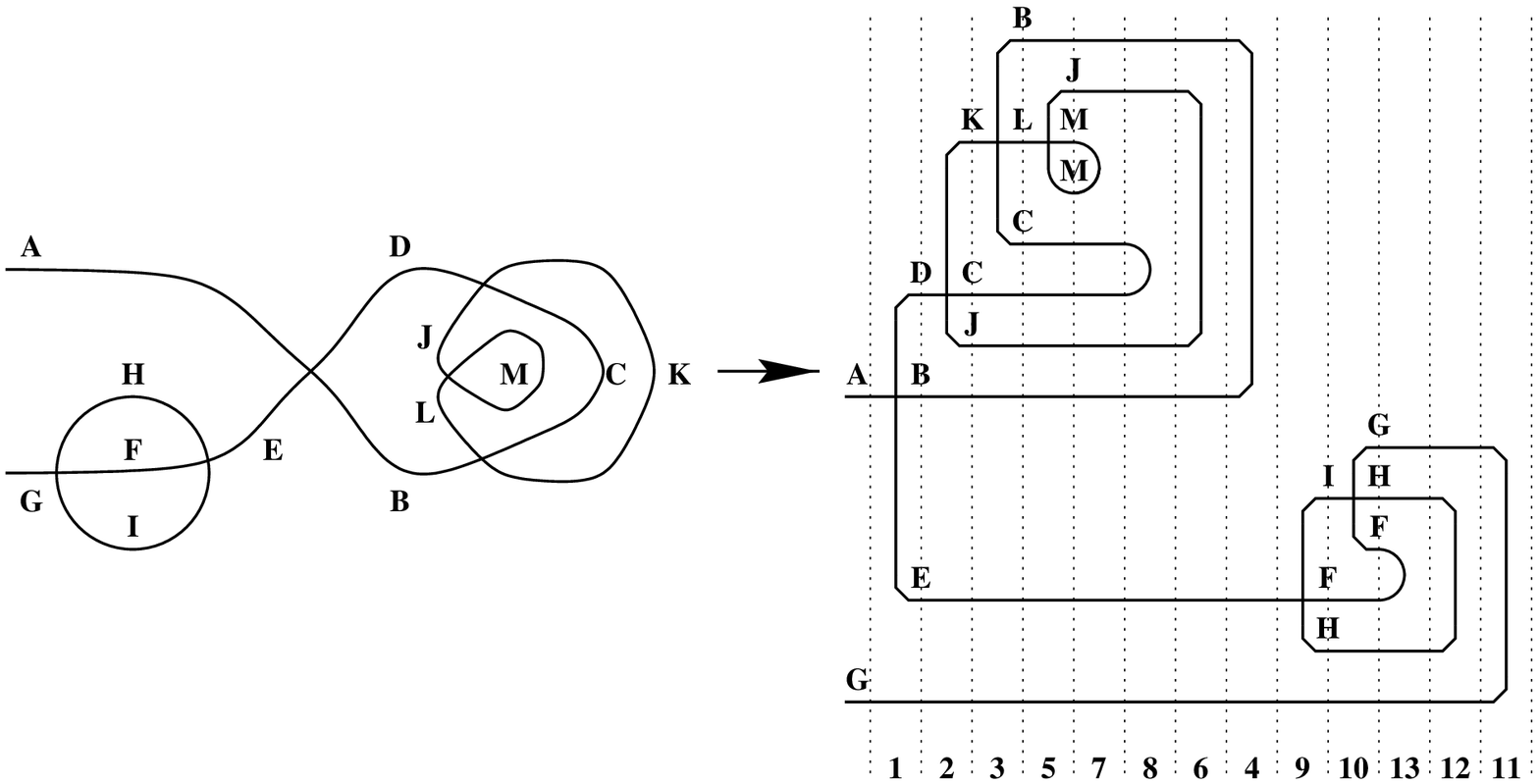}}

The single-step algorithm can be viewed as a generalization of this
principle. Let us, for the sake of illustration, consider the tangle
diagram shown in Fig.~\singlestep.
Since there is in general more than one connected component, clearly
the concept of a unique ``active line'' no longer applies.
Let us instead start from an initial state given by {\it both} external
legs (edges A and G).

Moving along either of the edges A or G, a new line segment
(DE resp.~HI) is encountered. The question then arises which of
these to process first. We resolve this ambiguity by stipulating
that {\it in any given state, we evolve the line which at that
instant is uppermost}.\foot{With the optimization to be discussed in
Sec.~4.2 we shall permit certain permutations of the lines. However,
the line being evolved is in all cases the uppermost in the given
{\it state}, though not necessarily in the corresponding time-slice
representation.}
At time $t=1$, the edge A thus becomes B, and the new line segment
DE is added. The edge D is now the new top line.

Analogously, at the instants $t=2$ and $t=3$, the top line (D resp.\ K)
crosses a new line segment, which is then added to the current state.
We can formalize this by stating the transformation rule shown
in Fig.~\transform.1. This generalizes the corresponding rule of
\JZJ, except that the ``active line'' is now replaced by the uppermost
line.

At time $t=3$, the new top line carries the label B, which
was however already produced by the transformation acting at $t=1$.
We therefore proceed, at $t=4$, to the identification of the two ``copies''
of B, joining them through an arch. This is an example of the general
transformation rule shown in Fig.~\transform.2. The addition of an
arch means that the lines intermediate between the two instances of B
(at positions $p=1$ and $q$ on Fig.~\transform.2) can henceforth not
communicate with the lines at the exterior of the arch. These
``trapped lines'' must therefore eventually evolve to the empty state
(vacuum), independently of the rest of the diagram. This observation
has two implications:
First, since both transformations conserve the parity (even/odd) of the
number of lines, $p-q$ must be odd. Second, due to the above-mentioned
rule that the current top line must always be treated first, the
evolution of a possible set of ``trapped lines'' must take place at
a later time. This means that when illustrating the sequence of moves
on Fig.~\singlestep, we cannot {\it draw} the time-slices in chronological
order. Note however, that the time ordering of the transformations is
given by the time labels shown in the bottom of Fig.~\singlestep.

\fig\transform{The two types of transformations in the single-step algorithm.
1) Addition of a new line segment. 2) Identification of the top line
(at position $p$) with another line (at position $q$), accompanied with
the creation of a new block. Several remarks apply to the relative positions
of $p$, $q$, $r$ and $s$ (see text).}
{\epsfxsize=7cm\epsfbox{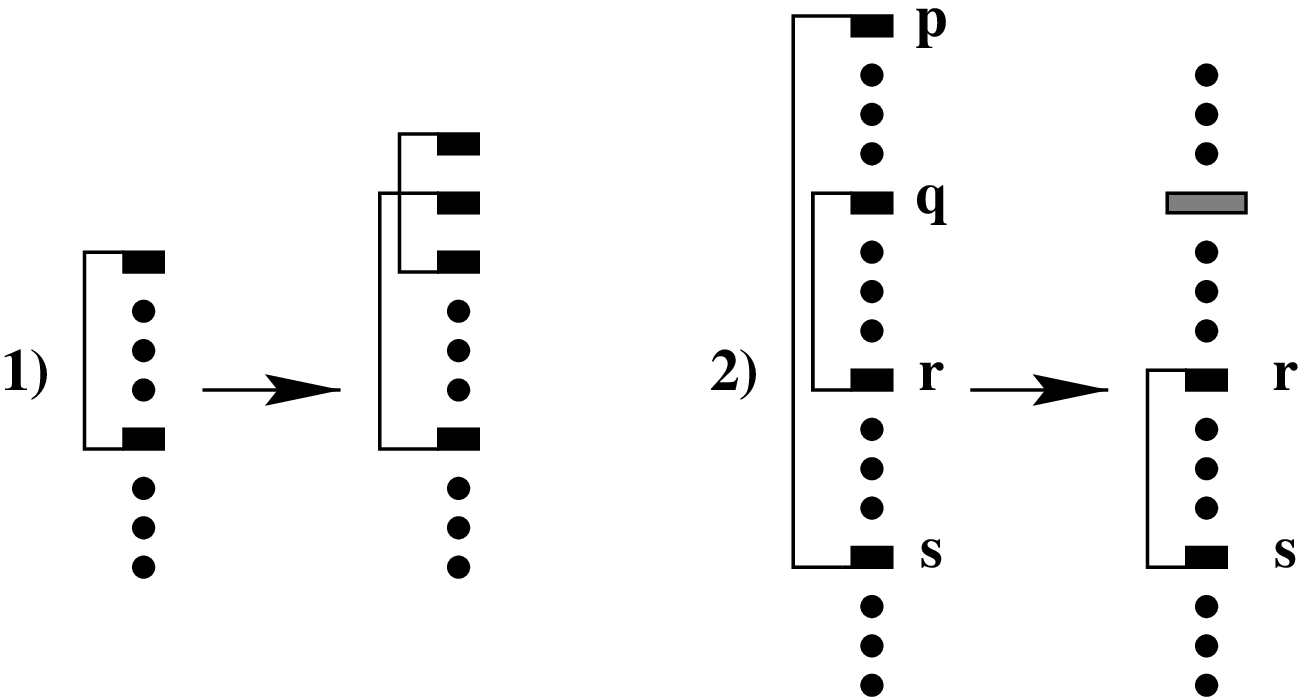}}

The existence of a number of trapped lines is visualized on
Fig.~\transform.2 by a {\it delimiter} (shown as a gray rectangle),
which seperates the remaining lines into two {\it blocks}.
Lines in different blocks cannot communicate, and must eventually
evolve to the vacuum separately. In particular this means that
the transformation 2) only applies when $p$ and $q$ belong to the
same block. Conversely, $q$ and $r$, and $r$ and $s$ may very well
be separated by one or more delimiters (not shown). Also, although
we have illustrated the case $r<s$, we may as well have $r>s$.

Of the thirteen transformations shown on Fig.~\singlestep, number
1, 2, 3, 5, 9, and 10 are of type 1, and the rest are of type 2.
Clearly, the two types of transformations increase (resp.~decrease)
the number of lines by unity. Since the initial state consists of
one line, the number of type 2 transformations must therefore
exceed the number of type 1 transformations by one.

The purpose of the transfer matrix is not only to count the total number
of tangle diagrams, but to do so for any fixed number of connected components.
In particular, when performing a type 2 transformation, we need to know
whether the points $p$ and $q$ were already connected though an arbitrary
number of edges at an {\it earlier} time. On Fig.~\transform\ we have
represented this information by a number of lines on the left, connecting
the points at a given instant into pairs. It may thus happen that on
Fig.~\transform.2, $r=p$ and $s=q$. In this case, the type 2 transformation
marks the completion of one connected component in the tangle diagram.

\fig\intermed{Intermediate states produced by applying the single-step
algorithm to the tangle diagram shown in Fig.~\singlestep.}
{\epsfxsize=10cm\epsfbox{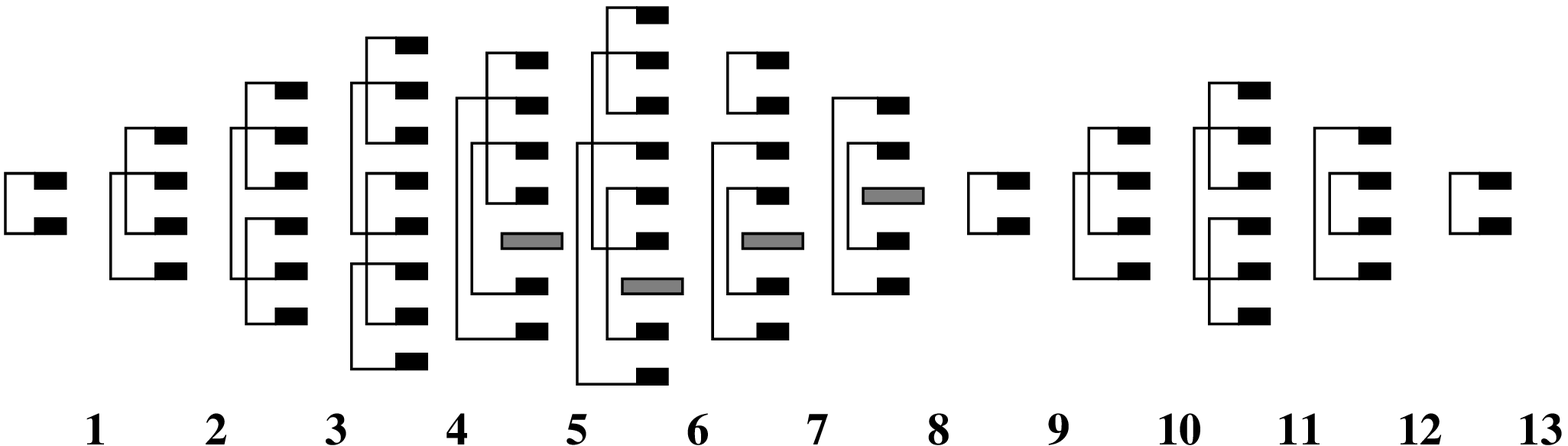}}

We are now ready to define the set of {\it states} on which the transfer
matrix acts. A state is defined by an even number of points (represented
on Fig.~\transform\ as black rectangles), connected into pairs by means
of the edges encountered at previous times. In addition, the points
are divided into $\ell+1$ blocks by means of $\ell \ge 0$ delimiters.
Note that points in different blocks can very well be connected, since
any connection made beforehand persists after the addition of a delimiter.

On Fig.~\intermed\ we show the set of intermediate states corresponding
to the time-slice representation of the tangle of Fig.~\singlestep.
The initial state of a two-legged tangle is given by a pair of points
(the exterior legs), implicitly connected at the point at infinity.
It can be noted that the same state may occur at different instants
of the transfer process. Also, any given state is not necessarily
allowed at all instants.

After each type 2 transformation one may be able to simplify the set
of delimiters. Namely, a delimiter may be eliminated if
it is adjacent to another delimiter, or if it precedes the first
point or succedes the last point of a state. On Fig.~\intermed\
we have implicitly assumed that such simplications have been carried out.

Finally, we must define the transfer matrix $T$ which counts {\it all}
tangle diagrams with $\tilde{p}$ vertices and $k$ connected components.
Its entries $T_{ab}$, where $a$ and $b$ are two basis states of the
kind just defined, are $0$ unless $b$ is a descendant of $a$.
An allowed state $b$ is a descendant of $a$ if it can be obtained
via a transformation of one of the two types shown on Fig.~\transform\ 
(for an arbitrary even $q\ge 2$ belonging to the same block as $p=1$),
followed by an arbitrary number of simplifications. $T_{ab}$ is
then the sum over all transformations from $a$ to $b$ of
the corresponding weight: $1$ or $n$ depending on whether
one closes a connected component or not ($n$ can be either
a given number, or a formal parameter, with a space of states
enlarged by polynomials in $n$ in the latter case).
For the moment the simplications are just the elimination of
superfluous delimiters. But we shall later (in Sec.~4.2) show that
hitherto different states are equivalent by means of suitable
transformations of the blocks, and of the points within each block,
such that a given state may be brought into a normal form. $T$ then
acts on the space of such normal forms. Apart from considerably
reducing the dimension of the state space, these additional simplifications
greatly enhance the efficiency of the algorithm.

As an example, we give in Appendix A the complete set of intermediate states
with their corresponding weights for the counting of tangle diagrams up to
four crossings.

\subsec{The geodesic algorithm}
We now turn to the description of our second algorithm.
Apart from providing a highly non-trivial check of our results, our
motivation for developing this alternative algorithm was to try to
limit the number of intermediate states and thus lower the memory
needs of the program. We still define the initial state as the set
of external legs, but we redefine the chronological order of the tangle
diagram by taking the time coordinate to be the
{\it geodesic distance to the set of external legs}.

Roughly speaking, in each time step we apply one of the transformations
shown in Fig.~\transform\ to {\it each} of the lines present in the
state at that instant. However, in order to introduce a valid time ordering
of the diagram this rough idea needs to be refined.

\fig\geodesic{The geodesic time-slice representation of the tangle diagram
given in Fig.~\singlestep.}
{\epsfxsize=4cm\epsfbox{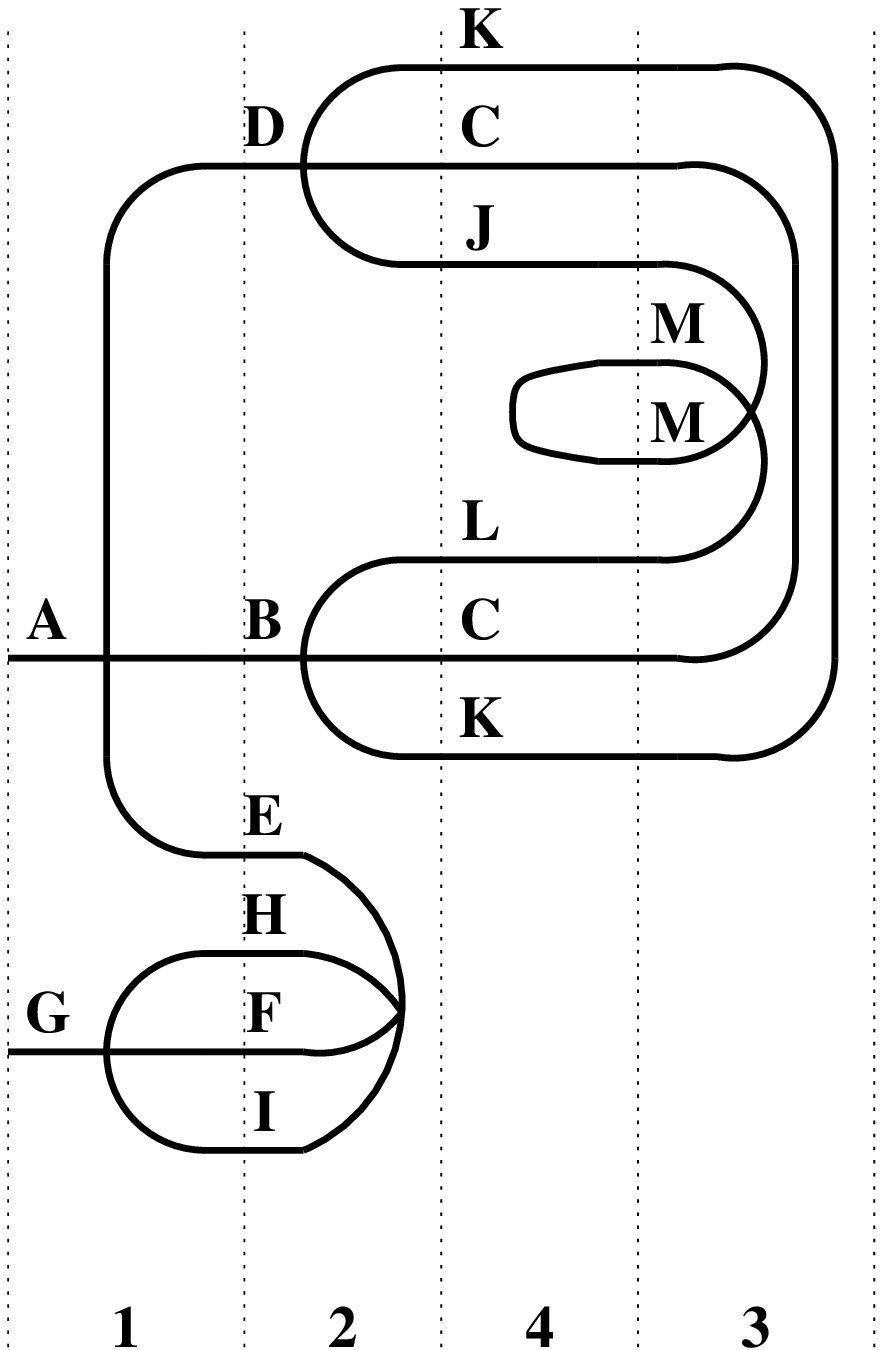}}

To progress, let us again consider the sample tangle diagram of
Fig.~\singlestep. In Fig.~\geodesic\ we show its new time-slice
representation, this time using the above geodesic definition of time.
At time $t=1$ the edges A and G are both subject to the same transformation,
in which three new edge labels (D, B, E resp.~H, F, I) are encountered:
as a short-hand notation we shall refer to this transformation as $1\to 3$.
It closely resembles the type 1 transformation in the single-step algorithm.
When $t=2$ the two upper edges (D and B) again undergo a $1\to 3$
transformation, whereas the four lower edges (E, H, F, and I)
annihilate at a common vertex: this is the $4\to 0$ transformation.
At the instant $t=3$, it is recognized that two edge pairs (K and C)
created at $t=2$ carry the same label and thus must be identified.
This transformation is reminiscent of the type 2 transformation in the
single-step algorithm, and we shall here tag it $2\to 0$.
At the same time, the two edges J and L cross so as to become a new pair
of edges, which are incidentally both labelled M. This is yet another
transformation, the $2\to 2$.

\fig\intergeo{Intermediate states used by the geodesic algorithm.}
{\epsfxsize=3cm\epsfbox{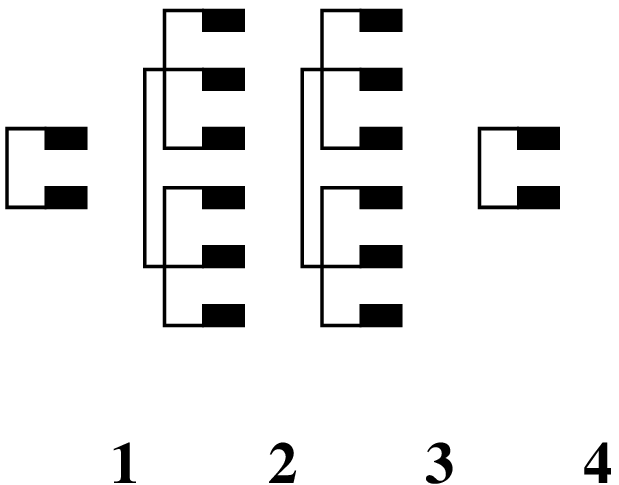}}

The intermediate states produced by this time-slice representation of
this example are listed in Fig.~\intergeo. When comparing with the single-step
algorithm (see Fig.~\intermed) we note a considerable simplification.

Turning now to the general case, we see that apart from the $2\to 0$
move, the transformations discussed above simply express the various ways of
``time ordering'' a tetravalent vertex, i.e.~to assign the label
$t$ to at least one of its incident edges, and the label $t+1$ to the
remaining edges. The possibility $0\to 4$ is excluded, as it would
lead to the creation of disconnected diagrams. This leaves us with
the transformations $4\to 0$, $3\to 1$, $2\to 2$ and $1\to 3$.
But due to the planarity of the diagrams, one also needs to take into
account that some of these transformations exist in several variants.
For example, the ``tadpole'' labelled M on Fig.~\geodesic\ is situated
to the left of its adjacent vertex; however, a different diagram exists
in which it is situated to the right. One must therefore accept that the
transformation $2\to 2$ comes in (at least) two guises:
in the first, the outgoing
edges bend backwards to the left, in the other they continue to the right.

\fig\geotrans{Schematic transformation rules for the geodesic transfer matrix.}
{\epsfxsize=5cm\epsfbox{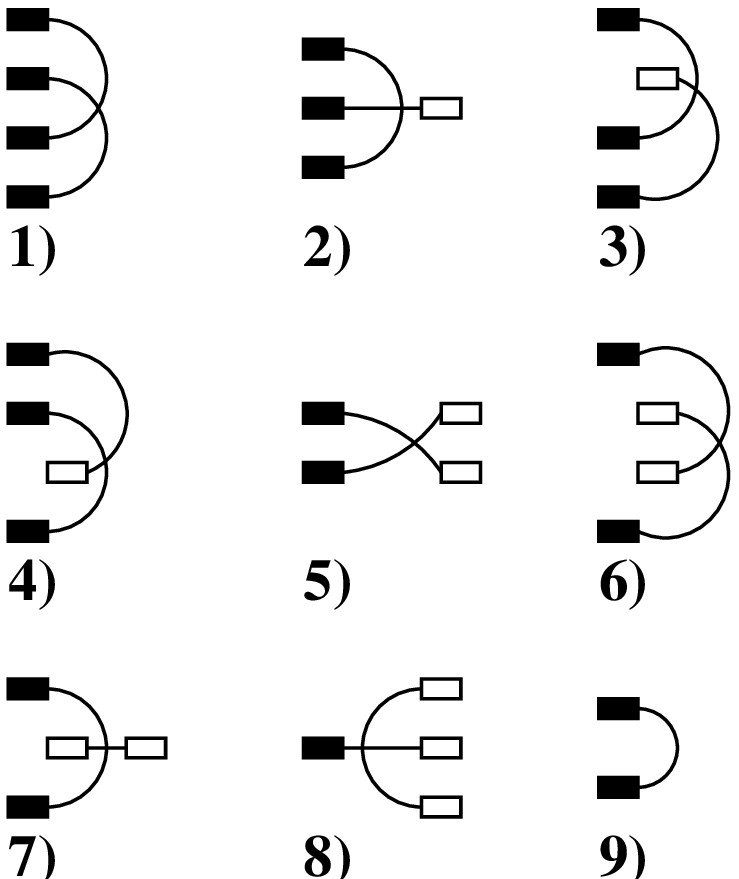}}

In Fig.~\geotrans\ we show schematically the complete set of transformation
rules. The black rectangles represent points at time $t$, and the white
ones points at $t+1$. The solid lines indicate the action of the transfer
matrix at time $t$. The first eight transformations are simply the 
topologically inequivalent ways of presenting a tetravalent vertex with
at least one point labelled by $t$. In particular we note that the
transformations of type $2\to 2$ and $3\to 1$ each occur in three different
variants. The more exotic possibilities 3, 4 
(resp.~5) start contributing to tangle diagrams with at least 4 (resp.~5)
crossings.
Finally, the ninth transformation is simply the $2\to 0$.

\fig\georules{Exact tranformation rules for the geodesic transfer matrix.}
{\epsfxsize=14cm\epsfbox{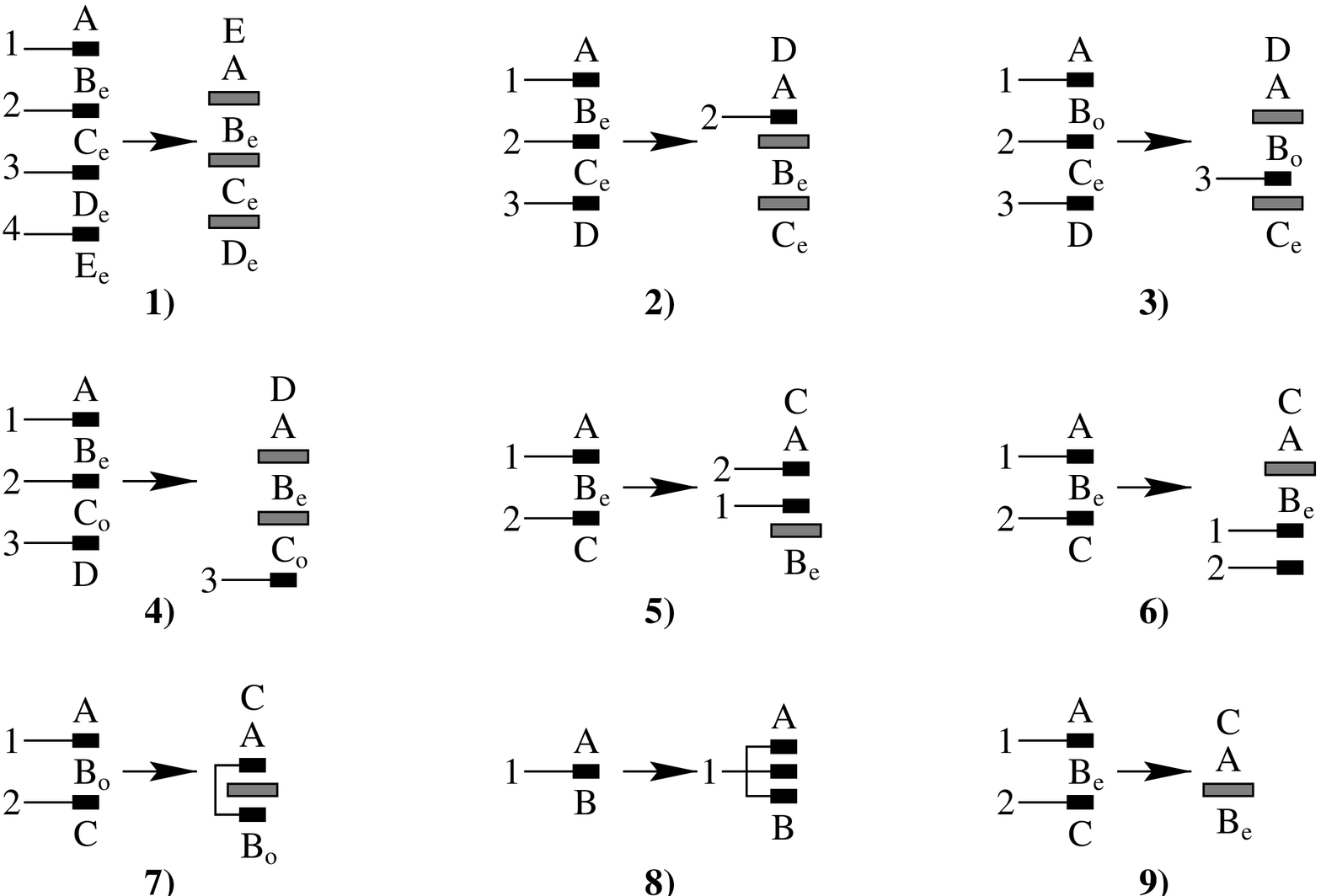}}

We still need to transcribe these rules in terms of the states previously
defined. In general, a given transformation leads to the creation of
several new enclosed regions (blocks), and to represent the latter in
terms of delimiters one needs to make use of the fact that cyclically
permuting the points within a given block yields a topologically
equivalent state (we shall come back to this point later, in Sec.~4.2).
Also, since each block must evolve separately to the
vacuum there are various parity constraints on the positions of the
points entering a given transformation.
By convention, we shall use capital letters to designate (possibly empty)
blocks of points. Subscripts $e$ and $o$ indicate that the
number of points must be even (resp.~odd). Blocks with no subscript may
have any parity: they are however subject to the global constraint that
the total number of points must be even.
Represented in this way, the exact transformation rules are given
in Fig.~\georules.

Just like in the single-step algorithm, a transformation may be followed
by an arbitrary number of simplifications (see Sec.~3.1).

\subsec{Tangencies}
Until now we have been discussing the enumeration of tangle diagrams
in which every vertex represents a crossing. However, to account for
the flype equivalence we need to enumerate more general diagrams with
$p_1$ intersections and $p_2$ tangencies, as discussed in Section 2.
Fortunately, this is a very simple extension of either of our algorithms.

Let us for simplicity consider the case of the single-step algorithm.
Adding a tangency rather than an intersection is obtained by modifying
the transformation if Fig.~\transform.1, so that the two points added
at time $t+1$ are both immediately above (resp.~immediately below)
the uppermost point at time $t$. Calling these variants respectively
transformation 1a and 1b, a knot diagram with intersections and tangencies
is then generated by acting on the initial state with a sequence
of transformations 1, 1a, 1b and 2.

We also need to add to the characterization of each state a variable that, at
any given time, specifies how many tangency transformations (type 1a or 1b)
were used prior to that instant. The desired diagrams are then generated
by sequences of $p_1$ transformations of type 1, $p_2$ of type 1a or 1b,
and $p_1+p_2+1$ transformations of type 2, so that no intermediate state
is empty.

Omitting the details, we notice that it is equally straightforward to
include tangencies in the geodesic algorithm by obvious modifications
of the eight first transformations of Fig.~\geotrans.

\subsec{More external legs}
Another extension of our algorithms consists in the enumeration
of diagrams with $2\ell$ external legs, $\ell > 1$. To this end we
simply start from an appropriate initial state comprising $\ell$
line segments, instead of just one, and we demand that the total
number of type 2 transformations exceed the total number of type 1
(i.e. 1, 1a, or 1b) transformations by $\ell$.

For $\ell>1$, such diagrams come in several types, corresponding to
the number of ways of pairwise connecting the set of external legs
at infinity. More precisely, given an ordered set of $2\ell$
points $X_\ell \equiv \{x_1,x_2,\ldots,x_{2\ell}\}$,
the number of types equals the number of ways to divide the set
$X_\ell$ into pairs, considered up to the action of the dihedral group
$D_{2\ell}$ on $X_\ell$.\foot{This also has a group-theoretic
interpretation, in terms of number of $O(n)\times D_{2\ell}$-invariants in
the tensor product of $2\ell$ fundamental representations of $O(n)$,
for generic $n$.} In particular, there are two types
of (four-legged)
tangles (see Fig.~\types), and five types of tangles with $2\ell=6$
external legs (see Section 5).
The general integer sequence $1,2,5,17,79,\ldots$ thus defined is
discussed in \Liskovets.

\newsec{Implementational details}
Although both of the algorithms described in the previous section
are operational (as the reader may verify by studying Appendix A),
we still need to give various details relative to their implementations
on a computer. In particular, it is not clear how states of the type
shown in Fig.~\intermed\ may be conveniently represented and manipulated.
We shall address this question in Section~4.1. 
Another important observation is that states which until now have
appeared to be different are in fact topologically equivalent. We shall
discuss this point in Section~4.2 and demonstrate how it can be used to
improve the efficiency of both algorithms.
Finally, we compare the performances of the two different algorithms
(single-step and geodesic) in Section~4.3.

\subsec{Representation of the states}
In order to render the information contained in states of the type shown
in Fig.~\intermed\ machine recognizable we shall represent each of them by
an ordered list of non-negative integers. The length of the list representing
a given state equals the number of points in the state plus the number of
delimiters, and the order of its elements is given simply by reading
the state from top to bottom.
Each delimiter is represented by the digit zero. The other
points each correspond to a positive integer, with the convention that
two points are connected if and only if they are represented by the same
integer. Clearly, this convention is not unique: for instance,
$(1,2,0,1,2)$ and $(13,4,0,13,4)$ both describe the same state.
To get rid of this ambiguity we shall stipulate that each consecutive
digit, starting from the left, be chosen as small as possible, consistent
with the above rules. Thus, $(1,2,0,1,2)$ is the unique normal form
of our sample state.

The sequence of states shown in Fig.~\intermed\ can then be transcribed
as follows:
\eqna\newstates
$$\eqalignno{
 (1,1) &\to (1,2,1,2) \to (1,2,1,3,2,3) \to (1,2,1,3,2,4,3,4) \to
 (1,2,3,1,0,3,2) \cr
 &\to (1,2,1,3,4,2,0,4,3) \to (1,1,2,3,0,3,2) \to (1,2,0,2,1) \to
 (1,1) \cr
 &\to (1,2,1,2) \to (1,2,1,3,2,3) \to (1,2,2,1) \to (1,1). \cr}$$

At a given stage in the transfer process we need to run through the
states present at time $t$, apply the transformation rules described
in Section 3.1--3.2, and produce the set of descendant states (the states at
time $t+1$) with their respective weight. The first time a given
descendant state is produced, it must be inserted in a suitable
data structure along with is weight. If subsequently the same state
is produced again as a descendant of another parent state, rather
than inserting it again we need to retrieve it in the data structure
and update its weight. In order for the algorithm to be efficient,
the operations of insertion and retrieval must be accomplished in
constant time (i.e.~in a time that does not depend on the number of
states accomodated by the data structure).

These demands are fulfilled by a standard data structure known as 
a {\it hash table} \Hash.
It relies on the fact that to each state $i$ we can assign a unique integer
$k_i \in \Z_+$ (the hash key), and devise a function
$f: \Z_+ \rightarrow \{0,1,2,\ldots,P-1\}$ (the hash function)
that distributes the set of $k_i$'s more-or-less uniformly on the
set $\{0,1,2,\ldots,P-1\}$. By inserting the states $i$ into an
array of noded lists indexed by $f(k_i)$, we can retrieve any given state
in a time proportional to the mean length of one of the pointer lists,
$t \propto N/P$, where $N$ is the total number of entries.
In practice we choose $P$ to be a large prime such that $N/P \sim 10$,
and we use the hash function $f(k) = k {\rm \ mod \ } P$.

A convenient key $k_i$ can be defined by concatenating the list of
``digits'' entering the normal form of the state $i$ into one large integer.
To find the minimum number of bits required to store one digit,
we remark that for the counting of tangles with $\ell$ intersections
the digits are all $\le \ell+1$. In the case at hand this means that
we need to use at least five bits per digit; in practice we have
however chosen to use eight bits, in order to profit from standard
routines for handling character strings.

\subsec{Equivalences between states}
As has already been mentioned, it is true for either of the two tangle
enumeration algorithms that some of the states generated at a given
stage in the transfer process are topologically equivalent.
Clearly, it is of the utmost interest to factor out as many topological
equivalences as possible from the state space, since the memory demands
as well as the time consumption of the algorithm are roughly proportional
to the number of states being treated.

A first such equivalence is due to the fact that any two different blocks
of points must evolve separately to the vacuum, without any mutual interaction.
The relative position of the blocks is thus immaterial.
The standard version of either algorithm (say, version 1) can thus be
ameliorated by introducing a standard order among the blocks before
inserting a given state in the hash table (version 2). We have done
so by simply sorting the blocks according to their size. In the special
case of the single-step algorithm it is advantageous to place the
smallest blocks at the top of the state, since such blocks will then
be evolved to the vacuum before touching any other block. The small
block being eliminated, the remainder of the state will be smaller
and can thus be processed more expeditiously.
For the geodesic algorithm, the choice betwen ascending and descending
ordering is irrelevant.

We take the convention of not changing
the relative order of two equally sized blocks.\foot{Inspecting
by hand some modestly sized systems reveals that changing the order of
equally sized blocks will only lead to a very small further gain.
We have nevertheless made various attempts of imposing a more unique
way of arranging the blocks, but since such transformations tend to
break down a certain regularity in the connectivities which is
imposed by the type 1 transformation, these attempts actually resulted
in a slight {\it increase} in the number of states.}
After the permutation of the blocks, the representation of the state
in terms of a list of integers is brought back to its normal form
(see Sec.~4.1).

In Table~\equivstat\ we illustrate the resulting decrease in the number
of states inserted in the hash table. It seems clear that not only is the
number of states much smaller, but it even grows with a smaller exponent.

\tab\equivstat{Maximal number of intermediate states in the transfer
process for three different versions (see text) of the single-step algorithm
for tangle diagrams with $p$ self-intersections and no tangencies.}
{\vbox{\offinterlineskip
\halign{\strut\hfil$#$\hfil\quad&\vrule#&&\quad\hfil$#$\hfil\crcr
p&&{\rm version\ 1}&{\rm version\ 2}&{\rm version\ 3}\cr
\omit&height2pt\cr
\noalign{\hrule}
\omit&height2pt\cr
  2 &&       4 &      4 &     4 \cr
  3 &&       7 &      7 &     6 \cr
  4 &&      24 &     16 &    14 \cr
  5 &&      67 &     45 &    24 \cr
  6 &&     226 &    110 &    49 \cr
  7 &&     735 &    313 &   106 \cr
  8 &&    2573 &    804 &   209 \cr
  9 &&    9340 &   2160 &   479 \cr
 10 &&   32790 &   6345 &  1078 \cr
 11 &&  128794 &  17074 &  2382 \cr
 12 &&  468757 &  45858 &  5929 \cr
 13 && 1933350 & 127751 & 13992 \cr
%14 &&     ??? & 380259 & 33683
}}}

Second, the states are equivalent upon cyclic rotations (eventually combined
with a reflection) of the points within any given block. Since, once
again, the blocks are independent, these dihedral transformations can
be performed independently within each block.
The ultimate way of implementing this equivalence would be the following:
before inserting a state in the hash table, subject it to all possible
dihedral transformations, and check whether any of the transformed states
is already present in the table.
Unfortunately, the number of transformations increases faster than
exponentially with the size of the state, and this exhaustive search
would quickly end up usurping the majority of the CPU time.

We have therefore opted for a less perfect but much faster
alternative (version 3).\foot{Comparing with the exhaustive method
applied to some modestly sized systems shows that, once again,
the number of topological equivalences not detected by the
``approximate'' method is negligible when compared to the number
of states which are in fact topologically distinct.}
Having sorted the blocks according to their size, we consider in turn
all possible dihedral transformations on the points in the first block,
keeping fixed the positions of the points in the other blocks.
After each transformation we bring the integer representation of the
state into its normal form. We then identify (one of) the normal form(s)
which lexicographically precedes all the others, and lock the points
of the first block into their corresponding positions. Leaving the
first block locked, we procede to apply the same procedure to the
second block. We continue this way until all blocks have been locked,
and only then the resulting representation of the state is inserted
into the hash table.

The gain of version 3 over version 2 is comparable to the gain of
version 2 over version 1, as witnessed by Table~\equivstat.
In the following we shall therefore exclusively understand version 3
when referring to any one of the two algorithms (single-step or geodesic).

\subsec{Comparing the two algorithms}
A first striking difference between the single-step and the geodesic
algorithm can be observed by comparing how their respective number
of intermediate states (and thus the memory needs) evolve as a function
of the ``time'' defined by the transfer process.

\fig\memss{Memory profile of the single-step algorithm (version 3,
cf.~Sec.~4.2). The curves represent knot diagrams with $p=1,2,\ldots,19$
crossings and no tangencies.}
{\epsfxsize=8cm\epsfbox{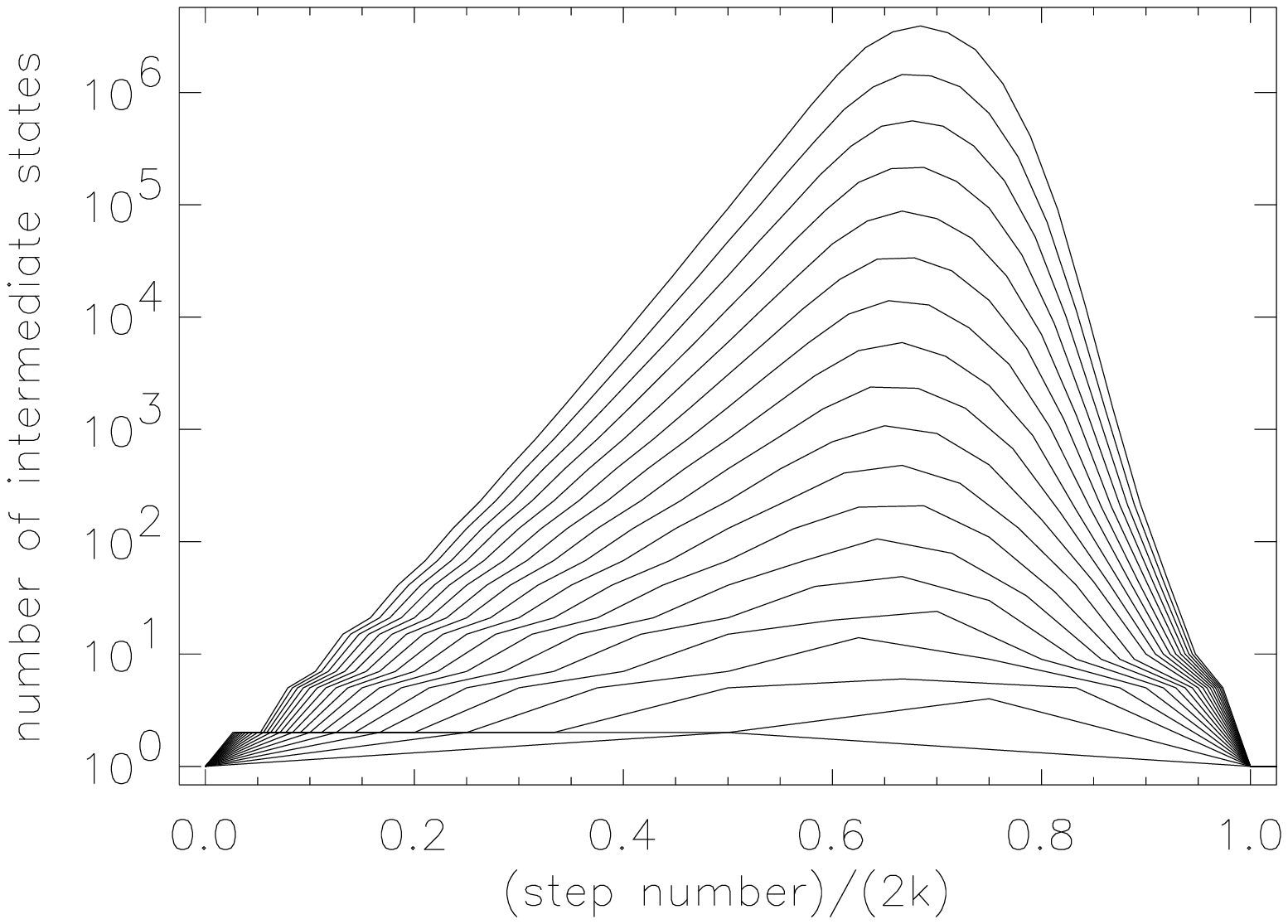}}

\fig\memss{Memory profile of the geodesic algorithm (version 2,
cf.~Sec.~4.2). The curves represent knot diagrams with $p=1,2,\ldots,10$
crossings and no tangencies.}
{\epsfxsize=8cm\epsfbox{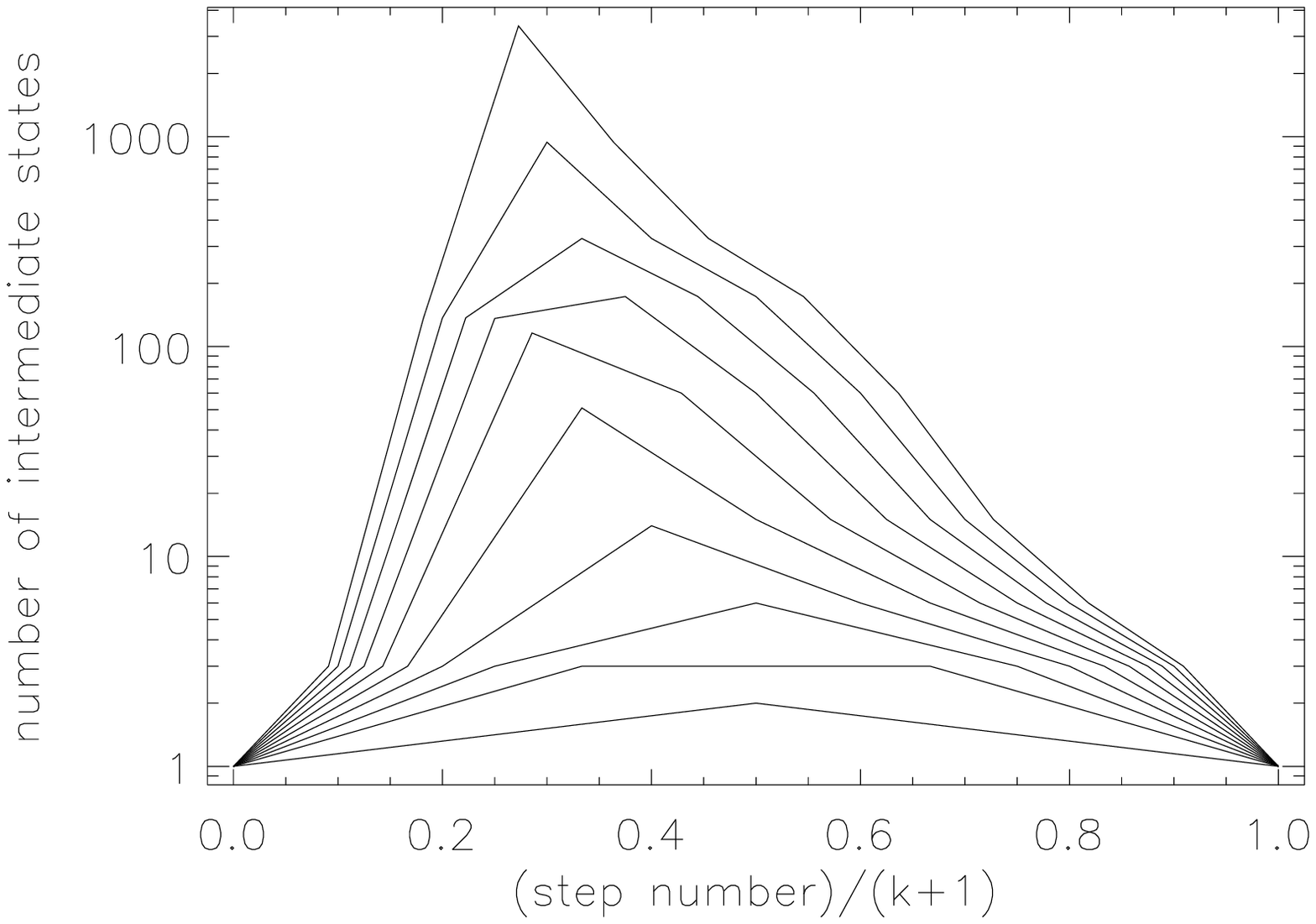}}

In both cases, the number of states grows exponentially in the beginning,
decreases exponentially towards the end, and reaches a maximum somewhere
in between. However, for the single-step algorithm this maximum is
reached at roughly $2T/3$ (where $T$ is the total number of time steps),
whereas for the geodesic algorithm the maximum is situated around $T/4$.
The reason for this difference is that the geodesic algorithm will
produce the majority of its states by applying the $1\to 3$ rule as
often as possible in the beginning of the process.

\tab\statgeo{Maximal number of intermediate states in the transfer
process for three different versions (see Sec.~4.2) of the geodesic algorithm
for tangle diagrams with $p$ self-intersections and no tangencies.}
{\vbox{\offinterlineskip
\halign{\strut\hfil$#$\hfil\quad&\vrule#&&\quad\hfil$#$\hfil\crcr
p&&{\rm version\ 1}&{\rm version\ 2}&{\rm version\ 3}\cr
\omit&height2pt\cr
\noalign{\hrule}
\omit&height2pt\cr
  2 &&    3 &    3 &    3  \cr
  3 &&    6 &    6 &    6  \cr
  4 &&   14 &   14 &   12  \cr
  5 &&   60 &   51 &   37  \cr
  6 &&  141 &  116 &   86  \cr
  7 &&  207 &  173 &  126  \cr
  8 &&  396 &  327 &  238  \cr
  9 && 1308 &  941 &  544  \cr
 10 && 5300 & 3367 & 1701  \cr
}}}

To estimate the actual memory needs of the algorithms, it is instructive
to compare the maximal number of intermediate states for the three
different versions defined in Sec.~4.2. For the single-step algorithm
the data were given in Tab.~\equivstat; we show the corresponding
numbers for the geodesic algorithm in Tab.~\statgeo. As expected, 
version 1 of the geodesic algorithm employs considerably fewer states
than version 1 of the single-step algorithm. However, quite surprisingly,
the ameliorations implied by version 2 and version 3 lead to an enormous
gain in the single-step case, but only a modest one in the geodesic
case. Thus, in version 3 the asymptotic growth of the number of states is
significantly slower in the single-step algorithm than in the geodesic one,
even though the latter was explicitly designed to use fewer states!
Although a qualitative explanation of this phenomenon can be given by
inquiring into the structure of a typical state we refrain from doing
this here. Rather, let us simply accept the efficiency of the single-step
algorithm as a remarkable fact.

Even when discarding the issue of memory, the geodesic algorithm
has a serious drawback compared with its single-step counterpart as far as
time consumption is concerned. Namely, the single-step algorithm
processes each state in a time that grows roughly linearly with its
size, whereas for the geodesic algorithm this time grows exponentially.
To see this, consider the intermediate state which is obtained from the
initial state by performing $p$ transformations of type $1\to 3$.
To turn this state into tangle diagrams with exactly $p$ crossings,
one needs to complete it with $p+1$ transformation of type $2\to 0$.
Supposing $p$ chosen so that the intermediate state has a complete
number of time slices, a total of $c_{p+1}$ diagrams will be
recursively generated,
where $c_k = {(2k)! \over k!(k+1)!}$ are the Catalan numbers.
The geodesic algorithm will therefore (asymptotically) spend the
majority of the CPU time closing up this ``maximally opened state''.

We have therefore used the geodesic algorithm as a highly non-trivial
check of our numerical results, but the data for large system sizes are
generated exclusively by the single-step algorithm.

To conclude this section, let us briefly discuss the time complexity
of our best algorithm (single-step, version 3). Based on the data
in Tab.~\equivstat, we infer that both time and memory needs
grow asymptotically as $\sim \kappa^p$, with
$\kappa \approx 2.7 \pm 0.2$.

\newsec{Numerical results}
We now present the numerical results that we obtained using the single-step
algorithm (version 3). Due to the enormous amount of data gathered we shall
only give the main results.

The first data are obtained by running a program that implements the single-step
algorithm without any tangencies. This corresponds to the generating function
$G(n,g_1=g,g_2=0)$ in the notation of Section 2. Its coefficients $a_{k,p,0}$
are given in Table \linkdiag\ up to $p=19$. The computation took
a few hours on a 1 GHz single-processor work station with 1 GByte of memory.
\font\four=cmr5 at 4pt
\def\sstrut{\hbox{\vrule height5pt depth3.5pt width0pt}}
\tab\linkdiag{Table of the number of alternating tangle diagrams with $2$ external legs.}{\vbox{\offinterlineskip
\halign{\sstrut{\four#}&\enskip\vrule#\enskip%
&&\hfil{\four#\ }\hfil\crcr
${}_{p}{}^{k}$&&0&1&2&3&4&5&6&7&8&9\cr
\omit&height2pt\cr
\noalign{\hrule}
\omit&height2pt\cr
0&&1\cr
1&&2\cr
2&&8&1\cr
3&&42&12\cr
4&&260&114&4\cr
5&&1796&1030&90\cr
6&&13396&9290&1349&22\cr
7&&105706&84840&17220&728\cr
8&&870772&787082&203568&14884&140\cr
9&&7420836&7415814&2312094&244908&6120\cr
10&&65004584&70867212&25691670&3575045&158354&969\cr
11&&582521748&685839770&282000444&48517524&3185314&52668\cr
12&&5320936416&6712285600&3074136464&628013796&55273668&1647728&7084\cr
13&&49402687392&66349573368&33387698708&7871666088&871779428&39142116&460460\cr
14&&465189744448&661680191832&361969672904&96451145091&12876308613&
786444610&16890227&53820\cr
15&&4434492302426&6651030871168&
3921901043440&1162484964230&181430681094&14126467392&462455640&
4071600\cr
16&&42731740126228&67329662060890&42499598861832&
13840075278704&2468480436152&234358127880&10552931952&171277860&
420732\cr
17&&415736458808868&685953949494774&460831546801414&
163246693686684&32699872694298&3666111325052&212581611050&5308497112&
36312408\cr
18&&4079436831493480&7028941367108708&5001468564165262&
1911737961254907&424232095742826&54835331971380&3912429396360&
135564649071&1722788176&3362260\cr
19&&40338413922226212&72403769391718890&
54341248085414380&22262254374655710&5413174461572394&791922013806504&
67266181855770&3025712334552&59605106568&326023280\cr}
}}
Let us first note that there are various quantities which can be extracted from
this table and which are known exactly in an independent way. They provide
a number of non-trivial checks. Let us define the number $a_p(n)$ of diagrams
at fixed number of colors $n$
\eqn\defcol{
a_p(n)=\sum_{k=0}^\infty a_{k,p,0} n^k
}
The simplest choice is to set $n=1$, that is to consider
the sum of each row of the table. This series of numbers is known exactly
(see \Tutte\ for a purely combinatorial argument, and \BIPZ\ for a field
theoretic one)
\eqn\exactoneb{
a_p(1)=2{(2p)!\over p!(p+2)!}3^p
}
and the corresponding generating function is
\eqna\exactone
$$\eqalignno{
G(1,g,0)&={1\over 3}A(4-A)&\exactone{a}\cr
A&={1-\sqrt{1-12g}\over 6g}&\exactone{b}\cr
}
$$

It is perhaps less well known that for $n=2$, one also has an exact
expression, in terms of elliptic integrals \KP:
\eqna\exacttwo
$$
\eqalignno{
G(2,g,0)&={1\over g^2}\left({g\over 2}-{1\over 8}{k\over(1+k)^2}\right)&\exacttwo{a}\cr
g&={E(k)-(1-k)K(k)\over 2\pi (1+k)}&\exacttwo{b}\cr
}
$$
where $K(k)$ and $E(k)$ are the complete elliptic integrals of the first
and second kinds.
We note that the generating function is non-algebraic,
and it is not known how to find it by direct combinatorial arguments.

There exists a similar, although more complicated,
formula for the case $n=-2$ which will be presented elsewhere \PZJc.

One can also find expressions for the last non-zero element of each row, which
formally corresponds to $n\to\infty$ (with $x\equiv n g^2$ fixed).
However there is a parity effect which
forces us to redefine separately odd and even generating functions:
\eqna\exactinf
$$
\eqalignno{
H_{\rm even}(x)&=\sum_{k=0}^\infty a_{k,2k,0}\, x^k&\exactinf{a}\cr
H_{\rm odd}(x)&=\sum_{k=0}^\infty a_{k,2k+1,0}\, x^k&\exactinf{b}\cr
}
$$
\fig\circles{a) Equation satisfied by $H_{\rm even}(x)$.
b) Recursive definition of a circle diagram
and its interpretation in terms of a rooted tree.}
{\epsfxsize=7cm\epsfbox{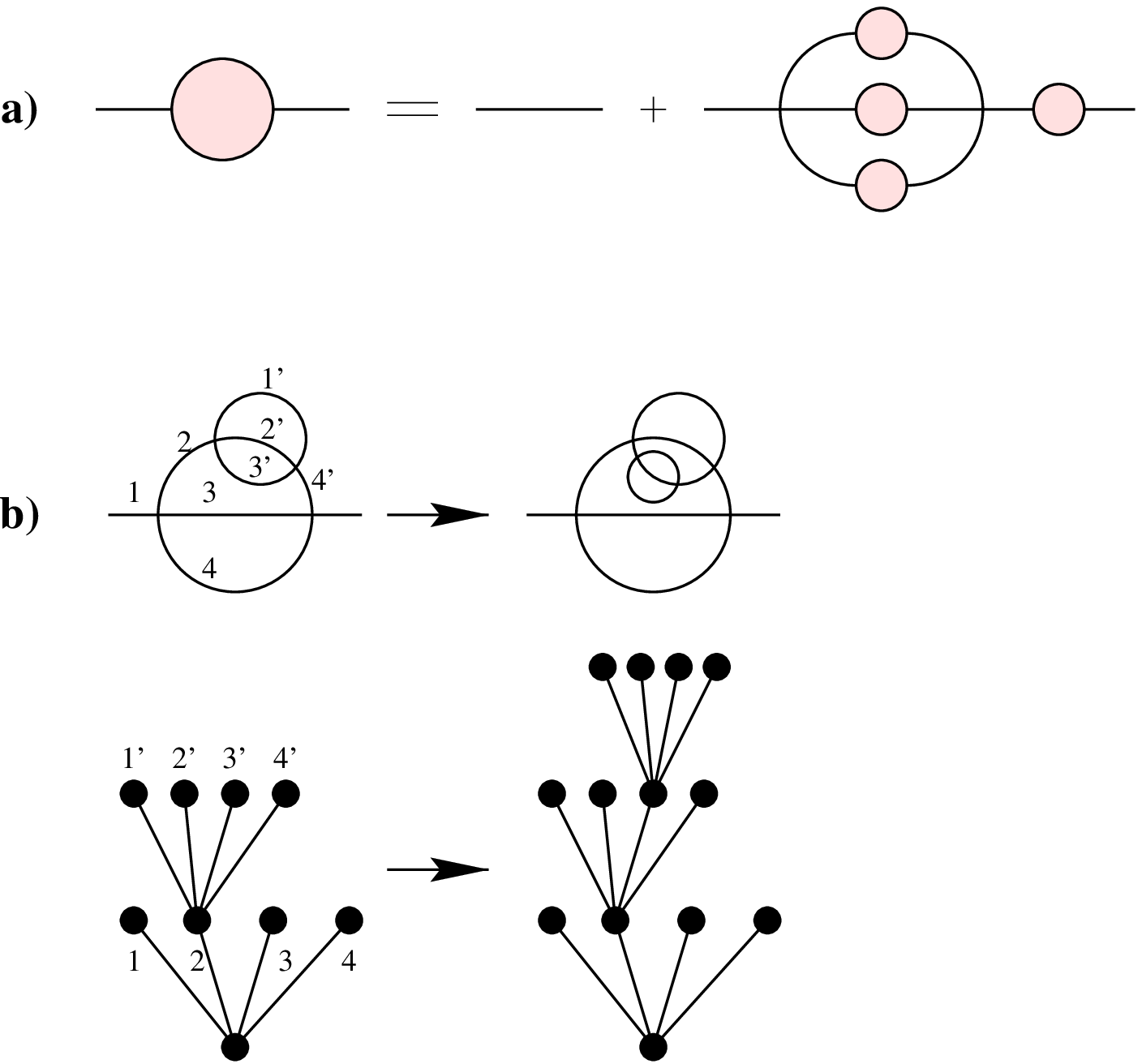}}

For the even case, a general diagram has the form of successive insertions
of circles in the bare propagator, which leads
to the equation, depicted on Fig.~\circles~a),
\eqn\exactinfbeven{
H_{\rm even}=1+x H_{\rm even}^4}
This can be described more explicitly in terms of rooted trees.
Each insertion of a circle requires two additional intersections, and it
leads to the creation of four new edges in which new circles can be inserted.
On Fig.~\circles~b) we have labelled two successive generations of edges as
$\{1,2,3,4\}$ and $\{1',2',3',4'\}$ respectively. Clearly, this reduces
the problem to that of enumerating rooted trees in which each node
(resp.~the root) can have degree 1 or 5 (resp.~0 or 4).
This is a simple example of a rather broad class of rooted trees discussed
by Tak\'acs \Takacs.

From Eq.~\exactinfbeven\ we infer that
\eqn\evencoef{
a_{k,2k,0} = {(4k)! \over (3k+1)! \, k!}
}

The proof of the formula for $H_{\rm odd}(x)$ is left as an exercize to the
reader:
\eqn\exactinfbodd{
H_{\rm odd}=2{\d\over\d x}(x H_{\rm even}^3)}

We infer that
\eqn\oddcoef{
a_{k,2k+1,0} = 2 {(4k+2)! \over (3k+2)! \, k!}
}

Finally, the first column in Table~\linkdiag\
reproduces the knot diagrams discussed in \JZJ, of course.

Next we want to deduce some properties of the asymptotic behavior of these series
from the numerical data.
The first quantity one can extract is the ``bulk entropy'' of
alternating tangles and links.\foot{The bulk behavior should be
independent of the number of external legs, and in particular be
the same for links and tangles.}
At fixed $n$ it is defined by the leading exponential behavior
of $a_p(n) \sim \e{\hat{s}(n)p}$:
\eqn\defent{
\hat{s}(n)=\lim_{p\to\infty} {\log a_p(n)\over p}
}
It would however be more natural to consider alternating links/tangles at
fixed number of connected components. This requires making an
appropriate scaling ansatz for the coefficients $a_{k,p,0}$, which
turns out to be
\eqn\defentb{
\log a_{k,p,0} {\buildrel k,p\to\infty \over \sim} p\, s(k/p)
}
where $s(x)$ is the bulk entropy at fixed ratio $x$ of the number of connected
components by the number of crossings.
It is clear from Eq.~\defcol\ that the two entropies defined above are related
to each other 
by a Legendre transform; namely, if one defines the {\it average} ratio $x$ at
fixed $n$:
\eqn\defentc{
x(n)={\d\over\d \log n} \hat{s}(n)=\left<{k\over n}\right>
}
then the following relation holds:
\eqn\defentd{
s(x(n))= \hat{s}(n) - x(n) \log n 
}
so that we also have the dual equation
of \defentc
\eqn\defente{
-\log n={\d\over\d x} s(x(n))
}

\fig\bulkt{The bulk entropy
$\hat{s}(n)$.}{\epsfxsize=8cm\epsfbox{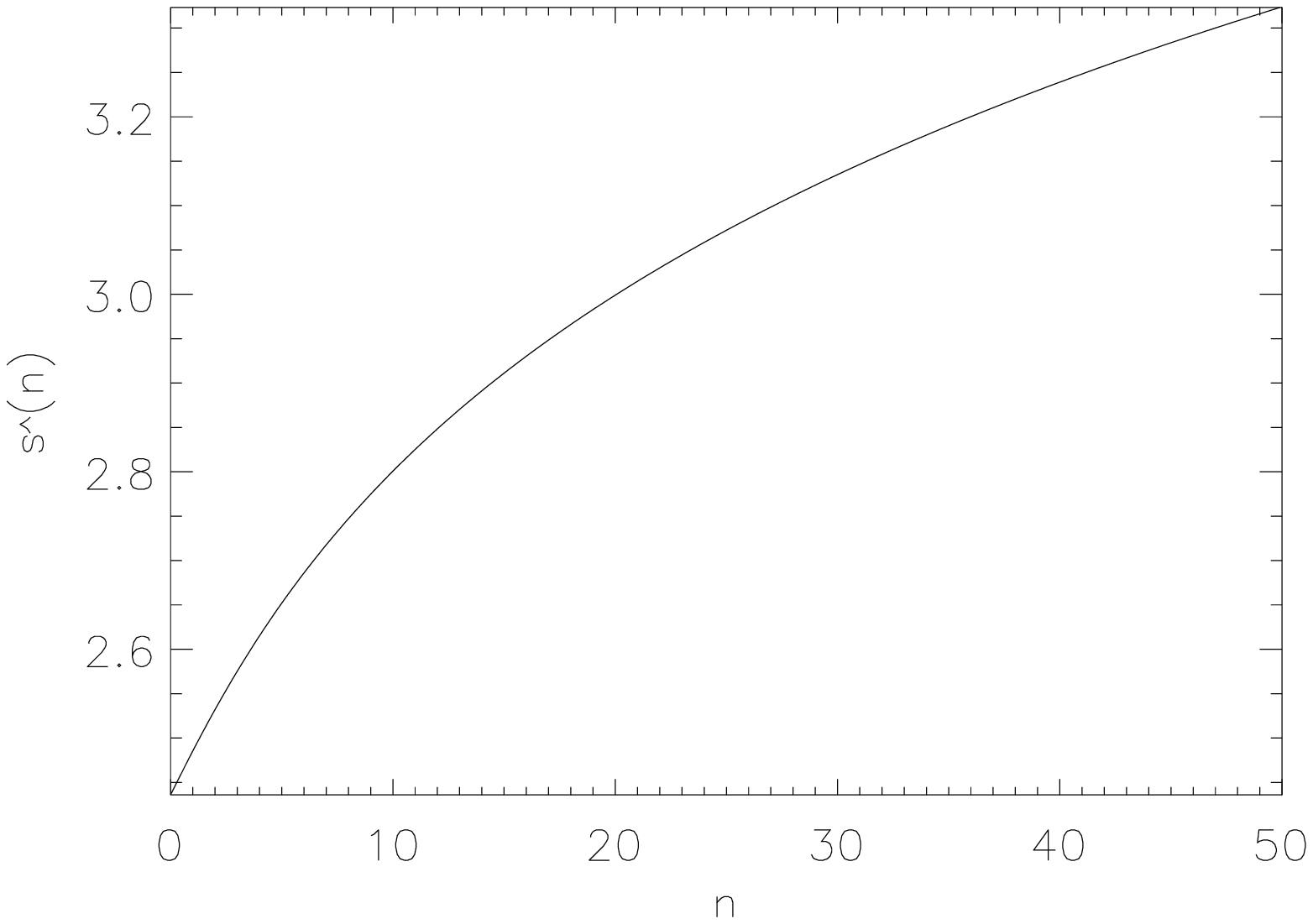}}
On Fig.~\bulkt\ the behavior of $\hat{s}(n)$ is shown for $n\in[0,\infty]$.
The various exact solutions mentioned above correspond to the following known
values of $\hat{s}(n)$:
\eqna\exacts
$$\eqalignno{
\hat{s}(1)&=\log 12\cr
\hat{s}(2)&=\log 4\pi\cr
\hat{s}(n)&{\buildrel n\to\infty\over =}{1\over2}\log n+\log{16\over 3\sqrt{3}}+o(1)\cr
}
$$
In \JZJ, the following numerical value was given: $\exp\hat{s}(0)\approx 11.42$,
which is confirmed here.

\fig\bulk{The bulk entropy $s(x)$.}{\epsfxsize=8cm\epsfbox{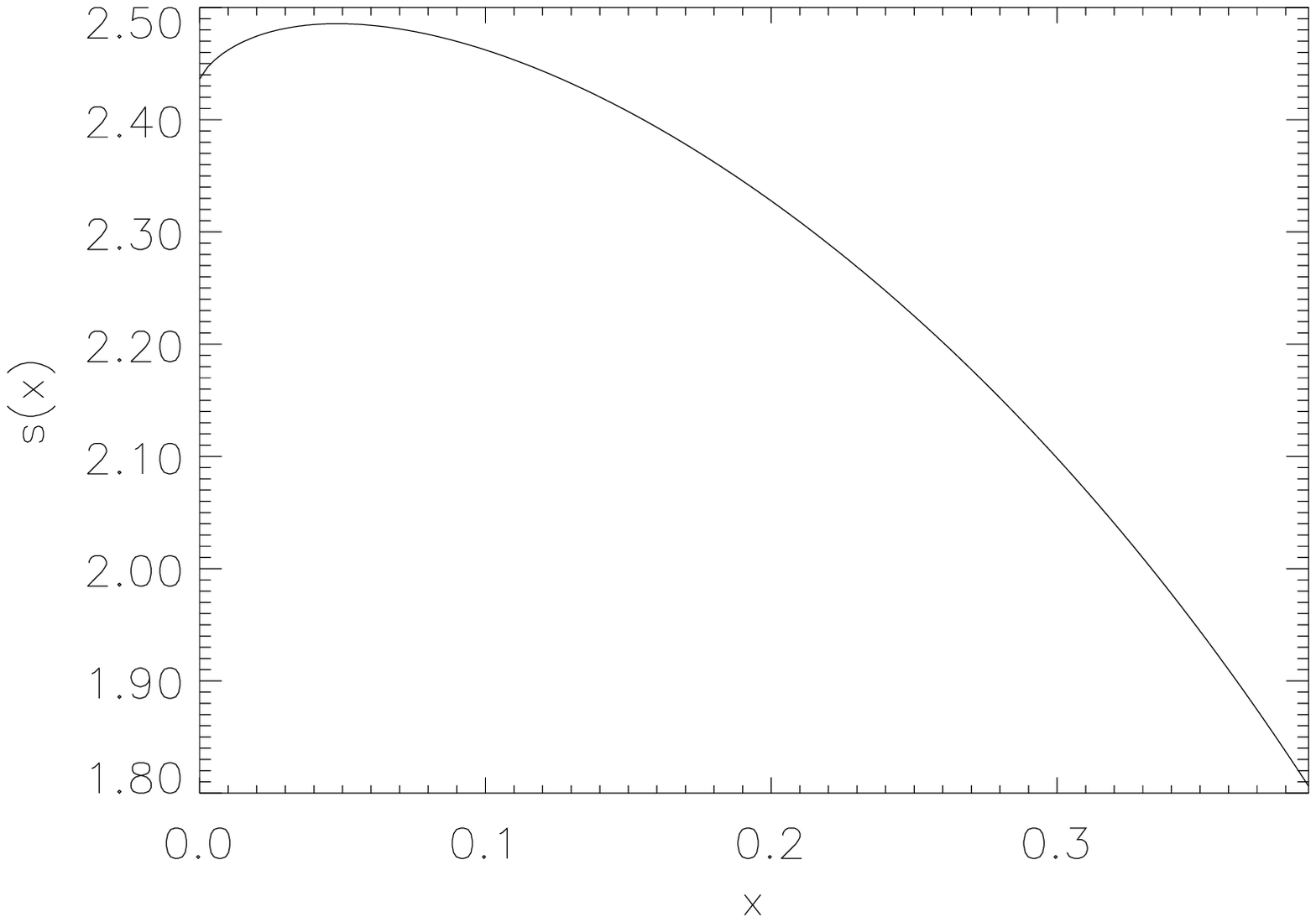}}
Even more interesting is the curve $s(x)$ shown on Fig.~\bulk\ for $x\in[0,1/2]$.
It displays a clear maximum at a value which is nothing but $x(1)$, cf
Eq.~\defente. Numerically we find:
\eqn\xc{
x(1)=0.0481\pm0.0001
}
This number has the following significance: at fixed number of crossings
$p$, a typical tangle/link diagram 
will have $0.0481\cdots \times p$ connected components when
$p$ goes to infinity. Note that this number is extremely small: any
average made over (equally weighted) alternating link diagrams
will be dominated by objects with few connected components.

Let us note that the bulk quantities {\it are affected} by the various 
renormalizations of Eq.~\ren{}, i.e.\ restriction to 2PI diagrams and inclusion
of the flype equivalence. However, it is expected that the qualitative properties
(and in particular the maximum of the entropy for a very small value of $x$)
are unchanged. For example, if one considers 2PI diagrams (which corresponds
to the counting of reduced prime alternating link/tangle diagrams), one finds a maximum
at $x_2(1)\approx 0.033$ instead.

The discussion of the exponent associated to the subdominant power-law
behavior of the series $a_p(n)$ (or, equivalently, $a_{k,p,0}$) is much
more involved. We define the critical exponent
\eqn\critexp{
\alpha(n)=\lim_{p\to\infty} {-\log a_p(n)+p\, \hat{s}(n)\over\log p}
}
Let us first recall the conjecture made in \PZJ, which relies on several
hypotheses: a) the asymptotic behavior of $a_p(n)$ is related to a singularity
of the corresponding generating function $G(n,g,0)$ which has the physical
meaning of singularity of 2D quantum gravity, i.e.\ large link diagrams 
behave as continuum random surfaces for which conformal field theory
techniques apply (KPZ formula \KPZ); b) the model describing link diagrams with
$n$ colors is in the same universality class as the usual $O(n)$ model of
dense loops \KoS, which relies on the assumption 
that there is no phase transition
in the generalized $O(n)$ matrix model. For $|n|< 2$, $n=-2 \cos(\pi\nu)$
($0<\nu< 1$), this implies that
\eqn\expconj{
\alpha(n=-2\cos(\pi\nu))=1+1/\nu
}
Let us now discuss separately various regions of $n$ and the corresponding
numerical analysis.

\noindent$\bullet$ For $n<0$, a difficulty arises in 
that coefficients $a_p(n)$ do not have a fixed sign. This implies
in particular that the dominant singularity of the generating function
$G(n,g,0)$ is not necessarily on the real positive axis, as would be implied
by hypothesis a) above. Numerically it seems that pairs of complex conjugated
singularities do occur and become dominant in a large region of $n$ which
includes at least part of the interval $n\in[-2,0[$, thus invalidating
conjecture \expconj\ in this region.
The analysis of such behavior is fairly involved and we leave it to future work.

\noindent$\bullet$ $0\le n< 1$: at $n=0$ it was suggested in \JZJ\ that
even though conjecture \expconj\ is correct ($\alpha(0)=3$), there might be
a logarithmic
correction which spoils the asymptotic behavior of the coefficients. For $n$ small, 
we expect several singularities extremely close to the dominant singularity
making any analysis difficult. Estimates of the critical exponent do not
contradict \expconj, but they have very low accuracy; for example,
\eqn\exa{
n={\sqrt{5}-1\over2}\qquad \nu=3/5 \qquad \alpha_{\rm conj}={8\over3}\qquad
\alpha_{\rm num}=2.6\pm 0.1
}

\noindent$\bullet$ $1\le n\le 2$: we can first extract from the exact solutions
(Eqs.~\exactoneb{}, \exactone{}\ and \exacttwo{}) the asymptotics
\eqna\asyonetwo
$$\eqalignno{
a_p(1)&\sim 12^p p^{-5/2} {\rm cst}&\asyonetwo{.1}\cr
a_p(2)&\sim (4\pi)^p p^{-2} (\log p)^{-2} {\rm cst}&\asyonetwo{.2}\cr
}
$$
They are of course compatible with \expconj; however we note a logarithmic
correction in \asyonetwo{.2} which comes from inverting the singularity of 
Eq.~\exacttwo{}: $g-g_c\sim (k-k_c)\log(k-k_c)$. At this point it becomes clear
that in order to remove the $(\log p)^{-2}$ factor,
one just needs to perform an appropriate
functional inversion on the generating series $G(2,g,0)$. Applying the same
procedure to the numerical data of $G(n,g,0)$ for $1\le n\le 2$, one can then
use standard convergence acceleration methods and obtain precise
estimates of the critical exponent. They are in good agreement with \expconj;
for example,
\eqn\exb{
n=\sqrt{2}\qquad \nu=3/4 \qquad \alpha_{\rm conj}={7\over3}\qquad
\alpha_{\rm num}=2.34\pm 0.01
}

\noindent$\bullet$ $n>2$: let us first recall that we have found
exact expressions at $n\to\infty$ for odd
and even coefficients separately (Eqs.~\evencoef\ and \oddcoef).
Asymptotically,
\eqna\asyinf
$$\eqalignno{
a_{k,2k,0} &\sim \left({256\over27}\right)^k k^{-3/2} {\rm cst}&\asyinf{a}\cr
a_{k,2k+1,0} &\sim \left({256\over27}\right)^k k^{-1/2} {\rm cst}&\asyinf{b}\cr
}
$$
i.e.\ the bulk terms are identical but the critical exponents are different.
This can be understood easily since in the odd case there is one ``defect'' in the sequence of circles which corresponds to marking one connected component, that is multiplying by $k$ (cf the differentiation in Eq.~\exactinfbodd).

Numerically, it is clear that for all $n>2$ odd and even series behave
differently. However once cannot perform any serious analysis on these 
series since they are too short. It may be that the exponents of
Eq.~\asyinf{} are preserved for any $n>2$, or finite values of $n$ might
smooth the difference between odd and even series; the data we possess
are unconclusive on this issue.

Let us end this analysis by noting that contrary to the bulk terms,
critical exponents are expected to be {\it independent} of the various
renormalizations of Eq.~\ren{}, due to universality arguments.

We now turn to the data obtained by inclusion of tangencies.
For reasons of conciseness, we here refrain from displaying the
three-dimensional array of coefficients $a_{k,p_1,p_2}$;
these are electronically available from the authors upon request.
Even the final results are fairly cumbersome to treat and display,
so that we only show the results for the
number of prime alternating tangles up to $p=15$ (even though they
can be easily obtained for $p$ up to $18$ or $19$, as in \JZJ,
on a work station, and probably a bit further using
larger computers).
\tab\resknot{Table of the number of prime alternating tangles.}{\vbox{\offinterlineskip
\halign{\strut#&\enskip\vrule#\enskip
&\hfil#\ \hfil&\hfil#\ \hfil&\hfil#\ \hfil&\hfil#\ \hfil&\hfil#\ \hfil&\hfil#\ \hfil&\hfil#\ \hfil
&\enskip\vrule#\enskip
&\hfil#\ \hfil&\hfil#\ \hfil&\hfil#\ \hfil&\hfil#\ \hfil&\hfil#\ \hfil&\hfil#\ \hfil&\hfil#\ \hfil
\crcr
&&&&$\Gam_1$&&&&&&&&$\Gam_2$\cr
${}_{p}{}^{k}$&&0&1&2&3&4&5&6&&0&1&2&3&4&5&6\cr
\omit&height2pt&&&&&&&&height2pt\cr
\noalign{\hrule}
\omit&height2pt&&&&&&&&height2pt\cr
1&&1&&&&&&&&0\cr
2&&0&&&&&&&&1\cr
3&&2&&&&&&&&1\cr
4&&2&&&&&&&&3&1\cr
5&&6&3&&&&&&&9&1\cr
6&&30&2&&&&&&&21&11&1\cr
7&&62&40&2&&&&&&101&32&1\cr
8&&382&106&2&&&&&&346&153&24&1\cr
9&&1338&548&83&2&&&&&1576&747&68&1\cr
10&&6216&2968&194&2&&&&&7040&3162&562&43&1\cr
11&&29656&11966&2160&124&2&&&&31556&17188&2671&121&1\cr
12&&131316&71422&9554&316&2&&&&153916&80490&15295&1484&69&1\cr
13&&669138&328376&58985&5189&184&2&&&724758&425381&87865&6991&194&1\cr
14&&3156172&1796974&347038&22454&478&2&&&3610768&2176099&471620&52231&3280&103&1\cr
15&&16032652&9298054&1864884&193658&10428&260&2&&17853814&11376072&2768255&308697&15431&290&1\cr
}
}}
These data satisfy once more various non-trivial checks,
including the comparison with the table in the appendix of \STh\ (for $n=1$),
of the Tables 1 and 2 in \ZJZb\ (for $n=2$), and Table 3 of \JZJ\ (for $n=0$).

\fig\typeone{Large $n$ expansion of tangles.}
{\epsfxsize=7cm\epsfbox{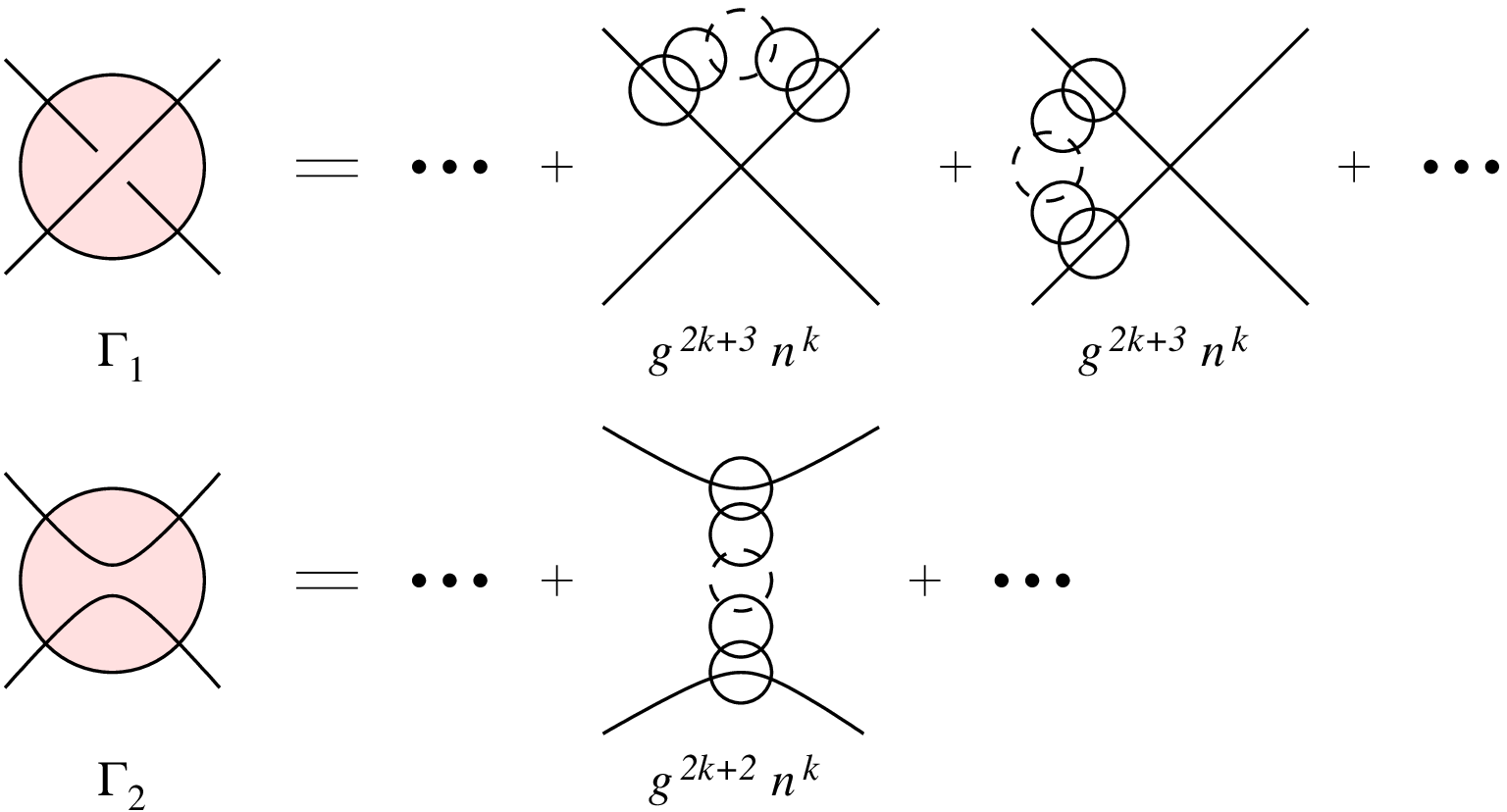}}
The last term in each column can also be verified by means of the large $n$
expansion. It is easy to convince oneself that
in general the 2PI diagrams with the largest possible number of connected
components is obtained by decorating the bare tangles by means of a festoon
of chained circles, as shown on Fig.~\typeone. In the case of $\Gam_1$
(resp.~$\Gam_2$) there are two (resp.~one) flype-inequivalent ways of
doing so.
As it stands, this argument holds true for odd $p$ (resp.~even $p$)
in the case of $\Gam_1$ (resp.~$\Gam_2$), but a similar reasoning holds
true for the opposite parity. We conclude that the last term in each row
of Tab.~\resknot\ should be $2$ (for $\Gam_1$ and $p \ge 6$)
resp.~1 (for $\Gam_2$ and $p \ge 2$), as is indeed observed.
The next to leading terms should be obtainable in a similar fashion.

Finally, we demonstrate the power of our method by applying it
to objects with more external legs. One may for example
ask how many ways there are to intertwine {\it three} strings,
and not just two as in the case of tangles. One must first
establish the different ways the strings are coming out, which leads
to Fig.~\typesix.
\fig\typesix{The five types of tangles with 6 external legs.}
{\epsfxsize=10cm\epsfbox{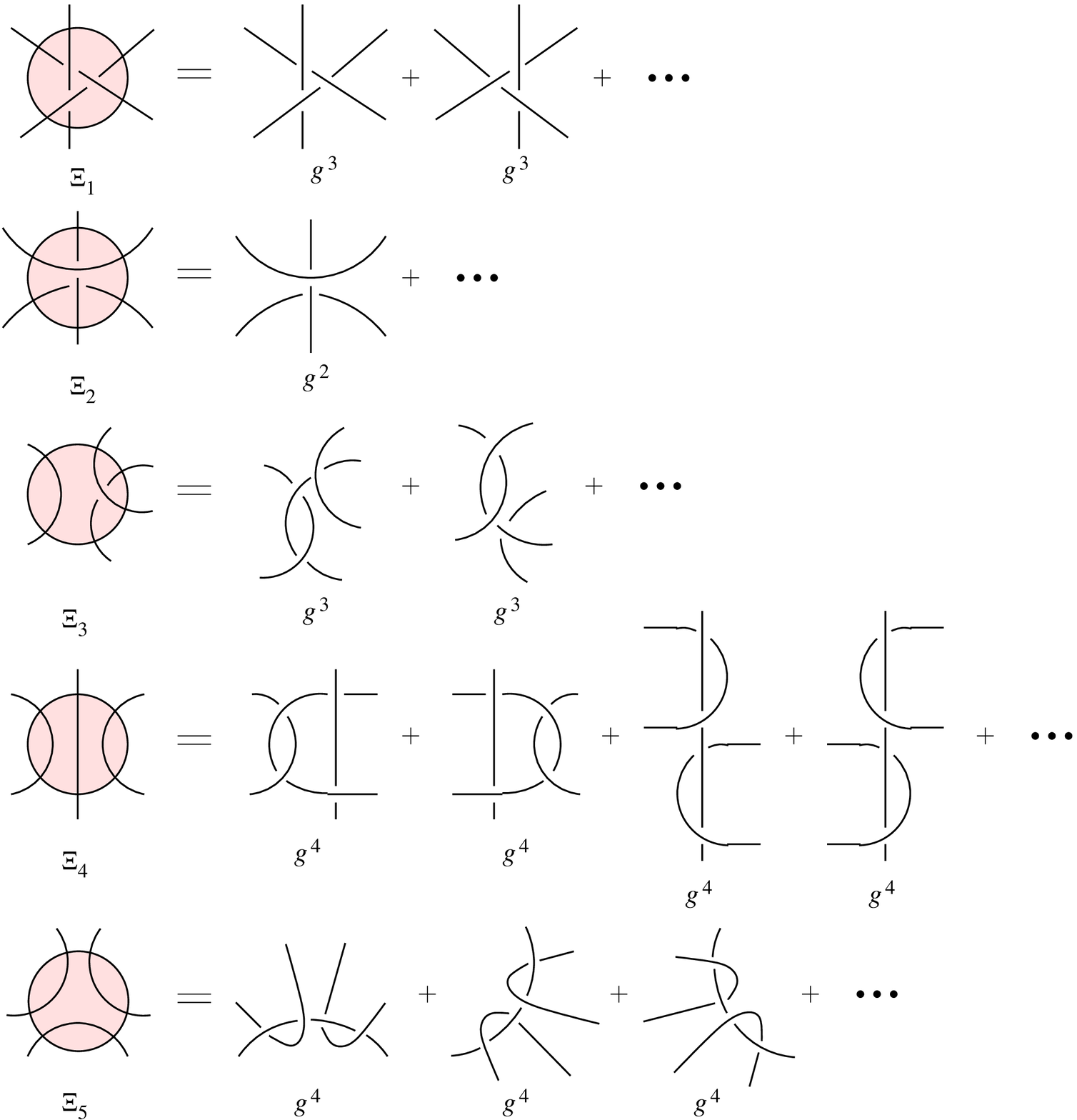}}

We only consider configurations such that no strings can be pulled out
altogether (``connected'' correlation functions in the language
of quantum field theory). Tab.~\ressix\ provides the first few
orders of the series of the numbers of such objects.
\font\seven=cmr7
\tab\ressix{Table of the number of prime alternating tangles with $6$ 
external legs.}{\vbox{\offinterlineskip
\halign{\strut#&\enskip\vrule#\enskip
&\hfil{\seven #\ }\hfil&\hfil{\seven #\ }\hfil&\hfil{\seven #\ }\hfil&\hfil{\seven #\ }\hfil
&\enskip\vrule#\enskip
&\hfil{\seven #\ }\hfil&\hfil{\seven #\ }\hfil&\hfil{\seven #\ }\hfil&\hfil{\seven #\ }\hfil
&\enskip\vrule#\enskip
&\hfil{\seven #\ }\hfil&\hfil{\seven #\ }\hfil&\hfil{\seven #\ }\hfil&\hfil{\seven #\ }\hfil&\hfil{\seven #\ }\hfil
&\enskip\vrule#\enskip
&\hfil{\seven #\ }\hfil&\hfil{\seven #\ }\hfil&\hfil{\seven #\ }\hfil&\hfil{\seven #\ }\hfil
&\enskip\vrule#\enskip
&\hfil{\seven #\ }\hfil&\hfil{\seven #\ }\hfil&\hfil{\seven #\ }\hfil&\hfil{\seven #\ }\hfil
\crcr
&&&$\Xi_1$&&&&&$\Xi_2$&&&&&$\Xi_3$&&&&&&$\Xi_4$&&&&&$\Xi_5$\cr
${}_{p}{}^{k}$&&0&1&2&3&&0&1&2&3&&0&1&2&3&4&&0&1&2&3&&0&1&2&3
\cr
\omit&height2pt&&&&&height2pt&&&&&height2pt&&&&&&height2pt&&&&&height2pt\cr
\noalign{\hrule}
\omit&height2pt&&&&&height2pt&&&&&height2pt&&&&&&height2pt&&&&&height2pt\cr
2&&0&&&&&1&&&&&0&&&&&&0&&&&&0&&&\cr
3&&2&&&&&0&&&&&2&&&&&&0&&&&&0&&&\cr
4&&0&&&&&7&&&&&2&&&&&&4&&&&&3&&&\cr
5&&18&&&&&6&&&&&16&2&&&&&8&&&&&9&&&\cr
6&&18&&&&&53&8&&&&42&2&&&&&42&7&&&&41&7&&\cr
7&&156&24&&&&154&6&&&&171&44&2&&&&156&14&&&&168&21&&\cr
8&&516&18&&&&609&181&6&&&748&114&2&&&&608&153&10&&&663&165&12&\cr
9&&2016&598&18&&&2956&422&6&&&2877&858&81&2&&&2850&586&20&&&3072&740&36&\cr
10&&10608&1428&18&&&11203&3498&318&6&&14037&3752&213&2&&&11918&3445&364&13&&13347&3966&438&18\cr
11&&40428&12318&1062&18&&57664&15330&738&6&&61028&19757&2511&131&2&&57602&17558&1406&26&&63393&20994&2040&54\cr
}}}

The lowest order in $p$ is explicited in Fig.~\typesix.
It is again relatively straightforward to check the correctness of the
last entry in each row of Tab.~\ressix\ by considering the $n\to\infty$
limit. In the cases of [$\Xi_1$ with $p \ge 8$],
[$\Xi_2$ with $p \ge 7$], and [$\Xi_3$ with $p \ge 3$] it is straightforward
to show that the diagrams having the highest power of $g$ are just
the trivial diagrams decorated by festoons (as in Fig.~\typeone),
meaning that the last entry in the corresponding rows should be
respectively 18, 6 and 2. For the cases
[$\Xi_4$ with $p \ge 4$] resp.~[$\Xi_5$ with $p \ge 4$] we conjecture that
the last entry of each row $\xi_p$ should read, for $p$ even,
\eqna\conj
$$\eqalignno{
\xi_p^{(4)} &= {1 \over 2}(3p-4)\cr
\xi_p^{(5)} &= {1 \over 8}(p+8)(p-2).\cr
}
$$
For $p$ odd we have $\xi_p^{(4)} = 2 \xi_{p-1}^{(4)}$
resp.~$\xi_p^{(5)} = 3 \xi_{p-1}^{(5)}$.

\newsec{Discussion and outlook}
In this paper we have shown how to efficiently enumerate alternating
tangle diagrams with a given number of connected components and
external legs, and we have explained how the flypes
can be easily incorporated into our algorithm to count only
topologically inequivalent objects. We have illustrated our method
with numerical data that are in agreement with the exact solutions
at $n=1$ \refs{\STh,\ZJZ,\PZJc}, $n=2$ \ZJZb, $n=-2$ \PZJc, 
and the limit $n\to\infty$, and that surpass the general results
of \ZJZb\ by several orders. From a computational point of view, we note
that the time required to compute order $p$ grows exponentially with $p$,
but much more slowly than the number of diagrams counted ($\sim 2.7^p$
compared to the number of diagrams $\sim 12^p$).

A remarkable advantage of our transfer matrix algorithms is that they
allow to generate planar diagrams with external legs, even in the
absence of a line or closed circuit defining an obvious transfer direction.

It should be noticed that our algorithms can be straightforwardly adapted
to graphs of any coordination number $q\ge 3$. For the single-step
algorithm, it suffices to modify the type 1 transformation so as to insert
$q-2$ new points, instead of just two. For odd $q$, the parity constraints
on transformation 2 no longer apply.%, and for two-legged diagrams the
%disconnected $1+1$ leg diagrams must be subtracted.

The geodesic algorithm can be similarly generalized. For example, on
Fig.~\geotrans, changing to trivalent vertices would simply imply having
five possible transformation rules instead of nine.

These generalizations open several interesting perspectives.
One obvious possibility would be to numerically study matter theories
defined on random graphs, by exact evaluation of correlation functions
\JZJc.

\vfill\eject
\appendix{A}{Tangle diagrams up to $4$ crossings.}
As an illustration we show on Fig.~\example\ the nine first iterations of the
transfer matrix. We restrict ourselves to states which generate diagrams with
at most four crossings. The states have all been simplified, as described
in Sec.~4. The weight of the trivial state (represented by a
cross on the figure) after step $2\ell-1$ gives the number of two-legged
tangle diagrams with exactly $\ell$ crossings, each connected component
being weighed by a factor of $n$.\foot{By convention, we also give a
weight $n$ to components connecting a pair of external legs, though
from the point of view of diagrammatic perturbation theory this is,
strictly speaking, not correct.}

\fig\example{The list of all intermediate states for tangle diagrams up to
four crossings. Below each state we indicate its weight.}
{\epsfxsize=12cm\epsfbox{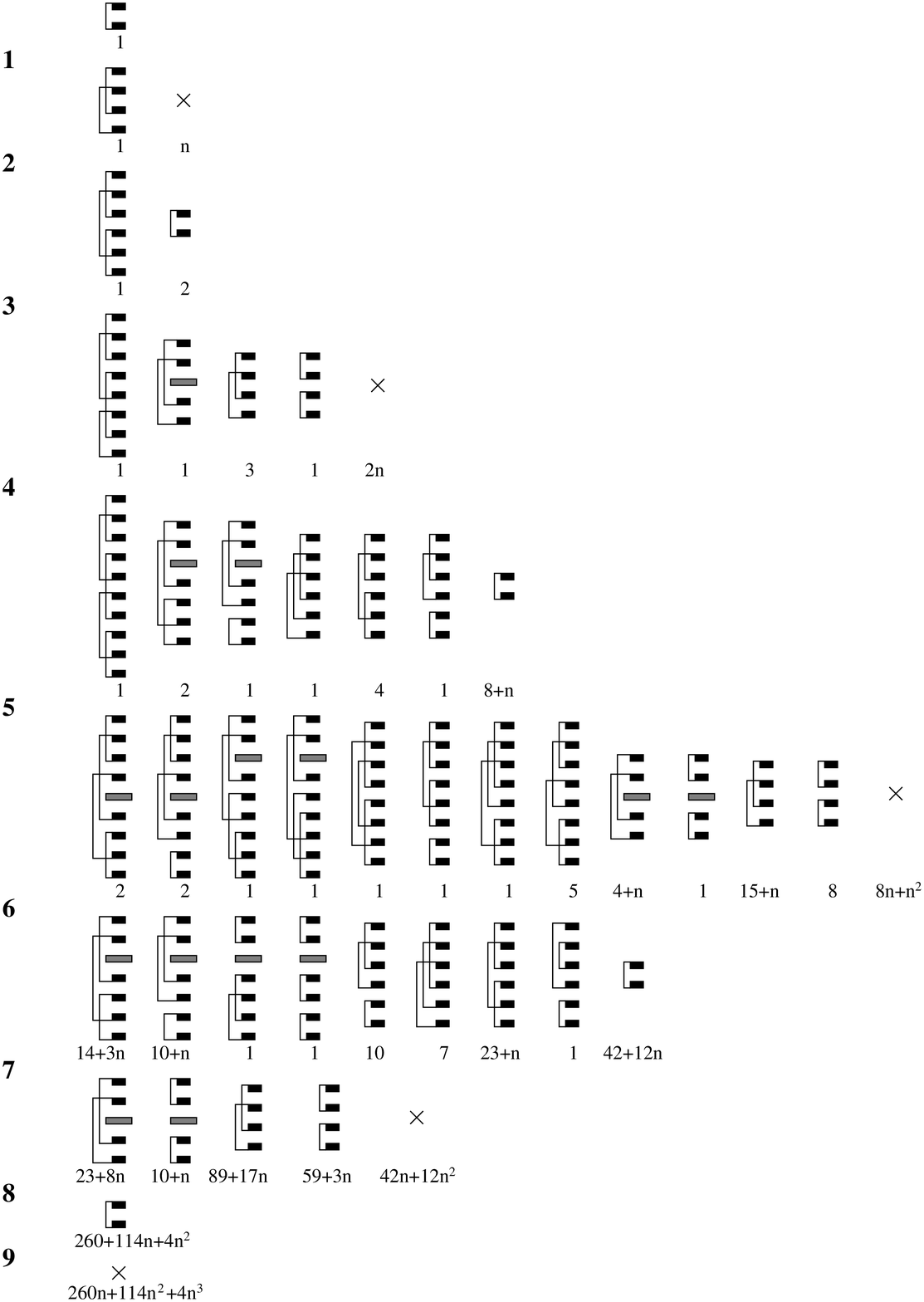}}

\footatend\vfill\supereject\immediate\closeout\rfile\writestoppt
\baselineskip=14pt\centerline{{\bf References}}\bigskip{\frenchspacing%
\parindent=20pt\escapechar=` \input refs.tmp\vfill\eject}\nonfrenchspacing
\bye